\documentclass[sigconf]{acmart}
\AtBeginDocument{%
  }

\copyrightyear{2027}
\acmYear{2027}
\setcopyright{cc}
\setcctype{by}
\acmConference[WSDM ’27]{Proceedings of the Twentieth ACM International Conference on Web Search and Data Mining}{February 15--19, 2027}{Hong Kong, Hong Kong}
\acmBooktitle{Proceedings of the Twentieth ACM International Conference on Web Search and Data Mining (WSDM ’27), February 15--19, 2027, Hong Kong, Hong Kong}
\acmDOI{10.1145/3862322.3863791}
\acmISBN{979-8-4007-3117-4/2027/02}

\usepackage{amsthm}
\usepackage{xspace}

\usepackage{enumitem}
\usepackage{multirow}
\usepackage{makecell}
\usepackage{subcaption}
\usepackage{bm}
\usepackage{mathtools}

\usepackage{algorithm}
\usepackage[noend]{algpseudocode}
\usepackage{pifont}

\newcommand{\defeq}{\vcentcolon=}

\usepackage{bbding}
\renewcommand{\qedsymbol}{\hfill$\blacksquare$}

\newcommand{\smallsection}[1]{\noindent\textbf{#1.}}

\newcommand{\mat}[1]{\mathbf{#1}}

\newcommand{\vect}[1]{\mathbf{#1}}

\newtheorem{problem}{Problem}

\DeclareMathOperator*{\argmax}{argmax}

\let\oldnl\nl
\newcommand{\nonl}{\renewcommand{\nl}{\let\nl\oldnl}}

\setlist[itemize]{topsep=1pt}
\setlist[enumerate]{topsep=1pt}
\newcommand{\arts}{\texttt{Arts}}
\newcommand{\movies}{\texttt{Movies}}
\newcommand{\video}{\texttt{Video}}
\newcommand{\sports}{\texttt{Sports}}
\newcommand{\industrial}{\texttt{Industrial}}
\newcommand{\beauty}{\texttt{Beauty}}

\newcommand{\Users}{\mathcal{U}}
\newcommand{\Items}{\mathcal{I}}
\newcommand{\Hist}[1]{\mathcal{H}^{(#1)}}

\newcommand{\method}{\textsc{CollabBank}\xspace}

\renewcommand{\algorithmicrequire}{\textbf{Input:}}
\renewcommand{\algorithmicensure}{\textbf{Output:}}

\newcommand{\AlgPhase}[2]{%
    \Statex
    \hspace*{\dimexpr-\labelwidth-\labelsep\relax}%
    \textbf{#1}:\hspace{0.2em}\textsf{#2}
}

\begin{document}

\title{\method: Compact Collaborative Memory for Accurate  Federated Recommendation}

\author{Daewon Gwak}
\affiliation{
  \institution{Soongsil University}
  \city{Seoul}
  \country{Republic of Korea}
}
\email{daewon1@soongsil.ac.kr}

\author{Jaehyun Park}
\affiliation{
  \institution{Soongsil University}
  \city{Seoul}
  \country{Republic of Korea}
}
\email{PjaeH@soongsil.ac.kr}

\author{Hanvit Lee}
\affiliation{
  \institution{Soongsil University}
  \city{Seoul}
  \country{Republic of Korea}
}
\email{hanvitlee@soongsil.ac.kr}

\author{Gyuwon Je}
\affiliation{
  \institution{Soongsil University}
  \city{Seoul}
  \country{Republic of Korea}
}
\email{gyuwonje@soongsil.ac.kr}

\author{Jinhong Jung}
\authornote{Corresponding author.}
\affiliation{
  \institution{Soongsil University}
  \city{Seoul}
  \country{Republic of Korea}
}
\email{jinhong@ssu.ac.kr}

\renewcommand{\shortauthors}{Gwak et al.}

\begin{abstract}
  Federated recommendation aims to build personalized recommendation models across distributed clients without collecting their raw interaction data on a central server. 
Effective federated collaboration requires sharing useful knowledge across heterogeneous clients. 
However, existing methods combine collaborative knowledge across clients without explicitly separating distinct behavioral patterns, potentially transferring irrelevant information while overlooking useful signals for a target client. 
Moreover, most learn such knowledge through iterative federated optimization, requiring repeated client–server rounds to update shared states.

In this work, we propose \method (Collaborative Memory Bank), a novel memory-based method that casts federated collaboration as retrieval over compact server-side memories. 
On the server, \method first uses semantically rich item embeddings from a pretrained language model to construct a small shared basis of semantic item groups and distills collaborative evidence from client groups sharing similar semantic patterns into separate memories stored in a memory bank. 
Afterwards, when a target client sends a compact query, the server retrieves and combines query-relevant memories from the bank and returns a client-specific memory that complements local recommendation.
Once the memory bank is constructed, each client retrieves its memory through a single round.
Through extensive experiments, we show that \method achieves strong accuracy against its competitors and validate the effectiveness of our collaborative memory design. 

\end{abstract}

\maketitle

\section{Introduction}
\label{sec:introduction}
\begin{figure}[t]    
    \centering
    \includegraphics[width=1\linewidth]{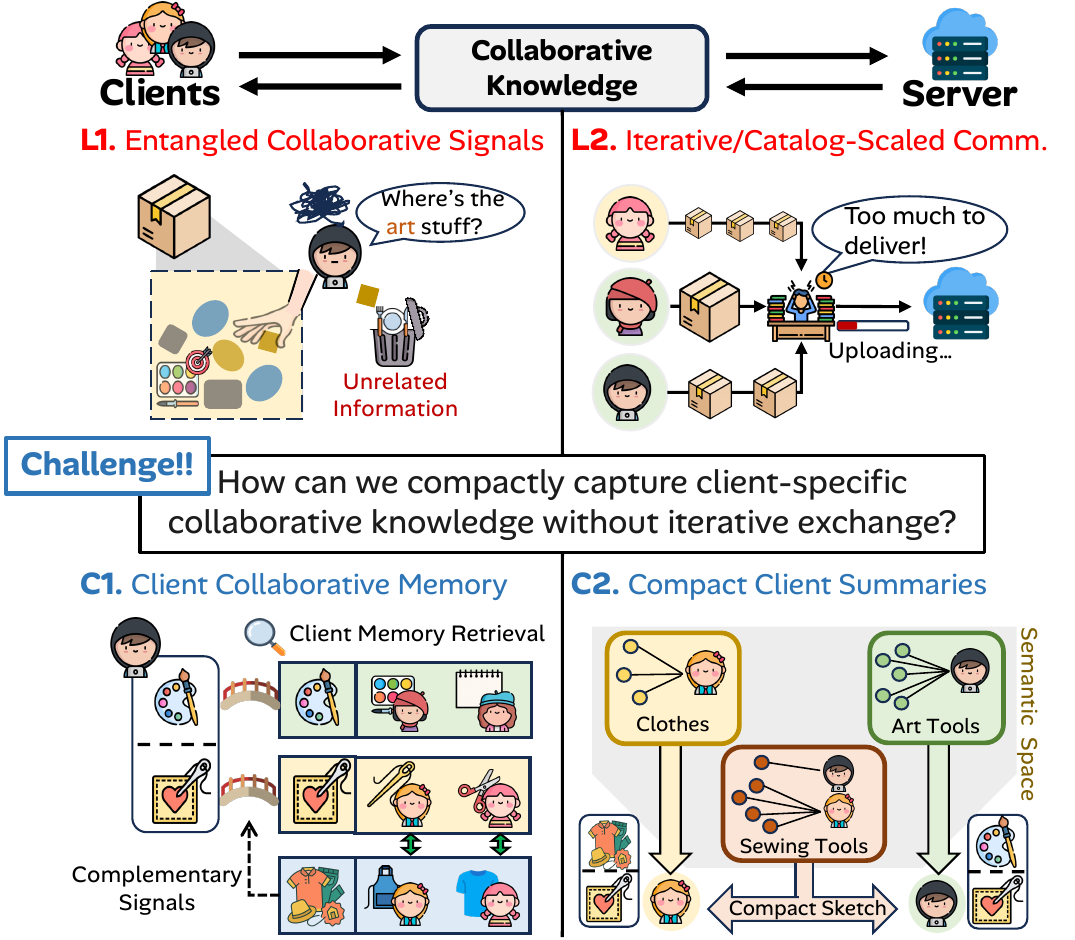}
    \caption{
        {Motivation of \method. It maps client behaviors to compact semantic groups and retrieves client-specific memories through non-iterative exchange to address entangled collaborative signals and costly communication.}
    }
    \label{fig:intro:teaser}
    \Description{The upper half contrasts entangled collaborative signals, including unrelated information, with repeated client--server uploads. The lower half depicts client memory retrieval and compact sketches over semantic groups such as clothes, art tools, and sewing tools.}
\end{figure}

{
Federated recommendation aims to collaboratively build personalized recommendation models across distributed clients without collecting their raw interaction data on a central server~\cite{LuoXS22,McMahanMRHA17,MintoHLH21,LinRCRY0RC20,10.1145/3799682.3840942}.
This problem is important because interaction histories, such as clicks, purchases, and ratings, reveal sensitive user preferences~\cite{ZhangWJ14,ParkJK17,LiuCCWNK22,abs-2502-11335,LiFYSDTJJLC26}, while transmitting them to a central server raises privacy risks~\cite{ZhangRWRCHZ21, NasrSH19, HuS24, BonawitzIKMMPRS17} and can be restricted by data-sharing regulations~\cite{3152676}. 
\leavevmode{For these reasons, federated recommendation has received considerable attention in data mining and has been explored for on-device personalization~\cite{ZhangL0ZYY24, LiL024, YuanYWZHW23, QuTZNHSY23, MaiDLDP25} and various real-world applications, including online content recommendation~\cite{QiWWH020, YiWWLS021, YuLWYC23, LiuCWLCC23}, healthcare~\cite{SongXZLYYO21,KwonKJJS24}, advertising~\cite{ZhangLTHM26, WeiDLTLWZ23,LeeKSJ24}, e-commerce~\cite{NiuWTHJLWC20, LiuXYFZM21}, and point-of-interest recommendation~\cite{ZhangYLZZY23}.}

Existing methods are broadly categorized according to how they represent collaborative knowledge. 
\textit{Model-based methods}~\cite{PerifanisE22, ZhangL0YZZY23, ZhangL0ZYY24, YuFLZWC26, HeLKH24} implicitly encode such knowledge in model states, such as model parameters, item embeddings, or gradients.
During each round of federated training, clients transmit their local model states to the server, which aggregates them into a shared state and sends it back to the clients~\cite{ChaiWCY21, PerifanisE22,ZhangYCHNC22,MuhammadWOTSHGL20}.
On top of that, prior work has explored additional client-side training to improve accuracy~\cite{ZhangL0YZZY23, YuFLZWC26, WangLHLCWL26, LiL00Z25}, along with resource-aware approaches designed to operate under client and system constraints~\cite{NguyenNNHLW24, LuJ0X025, YuanZHSW26}.
In contrast, \textit{relation-based methods}~\cite{NicolasSMCM25, Han0XLGZGL25} represent collaborative knowledge through structured item relationships, such as item-item similarity graphs, and propagate recommendation signals over these structures. 

{As illustrated in Figure~\ref{fig:intro:teaser}, existing methods have two limitations:}
\begin{itemize}[leftmargin=2em]
    \item[\textbf{L1.}]{
        \textbf{Entangled collaborative knowledge.} 
        They support global, personalized, or group-wise collaboration, but most represent and aggregate each client’s collaborative information as a whole. 
        Hence, signals supported by different portions of its interaction history and different client groups are merged into the same model state or relational structure. 
        This entanglement makes it difficult to exploit a client’s partial overlap with multiple groups and risks transferring unrelated information while overlooking useful patterns absent from its local history.
        }
    \item[\textbf{L2.}]{
        \textbf{Iterative or catalog-dependent collaborative exchange.} 
        Model-based methods update collaborative knowledge through iterative optimization, requiring repeated exchanges of client-updated model states. 
        Relation-based methods represent and exchange collaborative knowledge through item-level structures, making their payload grow with the item catalog (or the number of items). 
        {Thus, the cost of collaborative exchange scales with either optimization rounds or catalog size.}
    }
\end{itemize}

In this work, we propose \method (Collaborative Memory Bank), a memory-based method for accurate federated recommendation. 
To address these limitations, we cast federated collaboration as retrieval over a server-side bank of sparse cluster-level memories. 
Our key idea is to use semantically rich item embeddings from a pretrained language model to map heterogeneous local histories onto a small shared set of semantic item groups, providing a common coordinate system for compact client summaries and collaborative memories. 
From a one-time upload of these summaries, it builds separate memories from partially overlapping client subsets and stores them in the bank. 
When a target client sends a compact query, the server retrieves and combines query-relevant memories from the bank and returns a client-specific memory that provides signals complementary to local recommendation in a single retrieval round. 
Our main contributions are summarized as follows:
}
{ 
\begin{itemize}[leftmargin=1.2em]
    \item \textbf{Method.} {We propose \method, a memory-based method that maps local histories to a shared semantic basis, builds distinct memories from overlapping client subsets, and retrieves a client-specific memory for complement-aware recommendation.}

    \item \textbf{Communication.} We develop a non-iterative communication protocol that constructs the memory bank from one-time uploads of compact client summaries and supports single-round memory retrieval. After semantic initialization, its client-derived collaborative exchange depends on the compact cluster space rather than the number of items or optimization rounds.

    \item \textbf{Experiments.} Extensive experiments across six datasets demonstrate strong overall recommendation accuracy compared with federated baselines and consistent improvements over a local recommender using the same pretrained item embeddings. Additional analyses validate key design choices,  adaptation to evolving interactions, and  communication efficiency of \method.
\end{itemize}
}

\smallsection{Reproducibility} 
The code, datasets, and online appendix are available at
\url{https://github.com/daewongwak/CollabBank}.
 
\section{Related Work}

\smallsection{Model-based collaboration}
This approach encodes collaborative knowledge in model states through gradient-based optimization of learnable parameters and embeddings, with these states exchanged as payloads between clients and the server in each communication round.
Early approaches adapt conventional collaborative filtering models to the federated setting. FedMF~\cite{ChaiWCY21} keeps user representations local while collaboratively learning shared item representations, and FedNCF~\cite{PerifanisE22} extends this paradigm with neural interaction modeling. These methods establish the basic model-based collaboration paradigm, where clients repeatedly update local model states and the server aggregates them to capture cross-client collaborative knowledge.
Subsequent studies~\cite{ZhangL0YZZY23, ZhangL0ZYY24, HeLKH24, YuFLZWC26,ZhangLLZZL26, ZhangZLXXY26} improve this paradigm by enhancing model architectures or personalizing collaboration under client heterogeneity. For example, PFedRec~\cite{ZhangL0YZZY23} personalizes both user and item representations, GPFedRec~\cite{ZhangL0ZYY24} exploits client relationships to guide personalized collaboration, CoFedRec~\cite{HeLKH24} organizes clients into collaborative groups through co-clustering, and FedRecGEL~\cite{YuFLZWC26} improves the generalizability of shared item representations. {More recent studies also address resource constraints in federated recommendation~\cite{LinRCRY0RC20, NguyenNNHLW24, LuJ0X025, ZhangLWHL23}.}

\smallsection{Relation-based collaboration}
This approach represents collaborative knowledge as structured relationships between items and exchanges them as payloads~\cite{NicolasSMCM25,Han0XLGZGL25}.
For example, FedCIA~\cite{Han0XLGZGL25} computes an item--item similarity matrix from locally learned item representations, aggregates these matrices at the server into a global similarity structure, and uses it to align local item representations without directly averaging model parameters. 
This demonstrates that collaborative knowledge can be shared explicitly through inter-item relations while preserving client-specific representation spaces. 

Both paradigms entangle signals from different behaviors and client groups in model states or item-level structures, limiting behavior-specific transfer. Their communication cost also scales with training rounds or catalog size. \method maps local histories to a compact semantic cluster space and stores signals from partially overlapping client subsets in separate sparse memories. Retrieving and combining only query-relevant memories enables behavior-conditioned collaboration in one round, with communication cost independent of catalog size and training rounds.

In the broader literature, researchers have modeled positive and negative relations using signed graphs~\cite{ZhengWSWYZL25,ParkKJK25,GuJSJ25}, evolving interactions using dynamic graphs~\cite{LiZWLWY20,LeeJ23}, and higher-order interactions using hypergraphs~\cite{Xia0Z22,ChunLSJ24,GwakPKJ26}. These motivate a future extension of \method in which compact memories encode such relations for federated recommendation.

\label{sec:related_work}
 
\section{Preliminaries}
\leavevmode{We introduce basic concepts and define our research problem.}

\begin{figure*}[t!]
\centering
\includegraphics[width=1\linewidth]{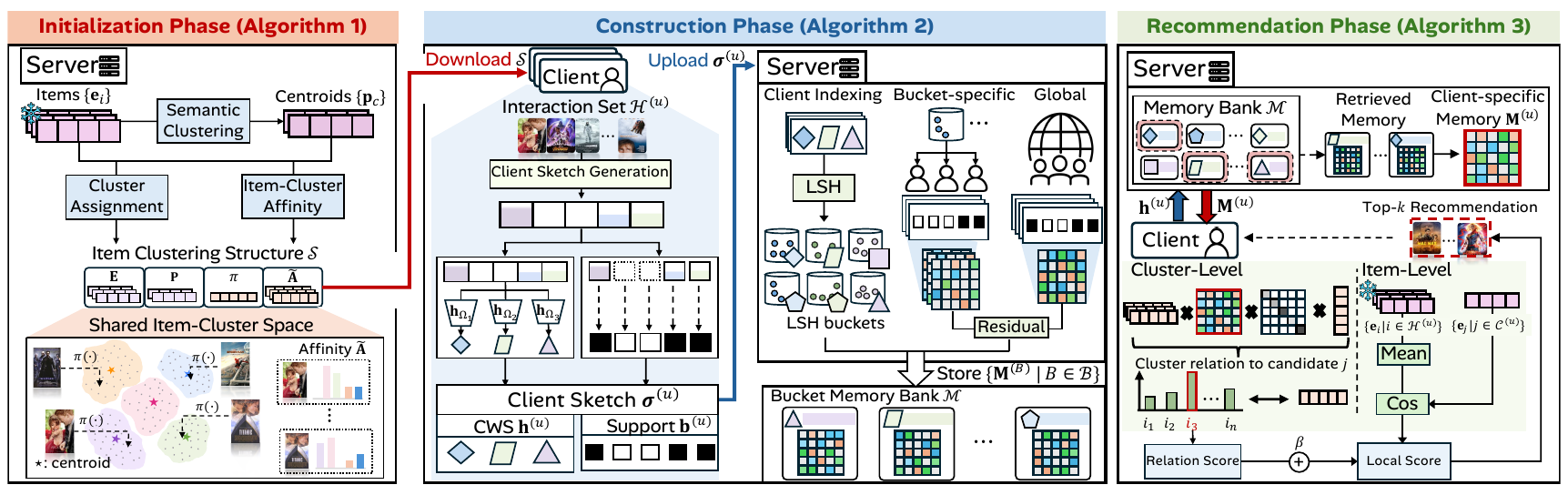}
\caption{
    Overview of \method. The initialization phase constructs a shared semantic item space. The construction phase builds the collaborative memory bank from compact client behavior sketches. The recommendation phase retrieves a client-specific collaborative memory and combines it with local scores to generate recommendations locally.
}
\label{fig:method:overview}
    \Description{Three panels show the data flow. The server clusters item embeddings and distributes the clustering structure. Clients summarize their histories as signatures and support vectors; the server uses LSH buckets and residual cluster associations to build a memory bank. A target client queries the bank, receives a client-specific memory, and combines relation and local scores to rank items.}
\end{figure*}

\subsection{Basic Concepts and Notations}
\smallsection{Vector and matrix}
For a vector $\vect{h}$, we denote $\vect{h}_i$ as its $i$-th entry. 
For a matrix $\mat{A}$, we denote $\mat{A}_{ij}$ and $\mat{a}_i$ as its $(i,j)$-th entry and $i$-th row, respectively. 
The lowercase symbol $\vect{a}$ is reserved for the rows of $\mat{A}$ and is not used for a standalone vector. 
We define $\mat{\tilde{A}}=\texttt{normalize}(\mat{A})$ as the row-normalized version of $\mat{A}$, where each nonzero row sums to one and each all-zero row remains zero.
We define $\texttt{sparsify}_R(\mat{A})$ as the row-wise sparsified version of $\mat{A}$, where at most the $R$ largest nonzero entries in each row are retained and all remaining entries are set to zero.

\smallsection{Notations for users and items}
Let $\Users$ and $\Items$ denote the sets of users and items, respectively.
We define $\Hist{u} \subseteq \Items$ as the interaction history of user $u$. 
Let $\mat{E} \in \mathbb{R}^{|\mathcal{I}| \times d}$ denote the matrix of precomputed {\color{black}L2-normalized item language representations (LRs),} where $\mathbf{e}_{i}$ is the $d$-dimensional embedding of item $i$ that encodes its metadata $t_i$, such as its title, using a pretrained language model (see more in the online appendix). 

\smallsection{Federated system}
We consider a federated system $\mathcal{F}$ organized in a
client--server architecture, comprising a server and
$|\Users|$ clients. 
Each client corresponds to a user and is connected to the  server.
Each client locally stores its interaction history $\Hist{u}$.

\subsection{Problem Definition}
{\color{black}We formally define the problem addressed in this work as follows:}
\begin{problem}[Federated Recommendation~\cite{PerifanisE22, ZhangL0YZZY23, HeLKH24}] 
\label{pre:problem_def} 
In the federated system $\mathcal{F}$, the goal is to generate personalized recommendations $L^{(u)}$ for each client $u$ by using collaborative information distributed across clients under the following setting:
\begin{itemize}[leftmargin=1.2em]
    \item \textbf{Local recommendation.}
    Each client $u$ locally generates $L^{(u)}$ containing the top-$K$ unseen items with the highest  scores.
    \item \textbf{Server coordination.}
    The server coordinates collaboration across clients \& aggregates their information, but does not generate $L^{(u)}$.
    \item \textbf{Raw-interaction locality.}
    Each history $\Hist{u}$ remains only on its corresponding client and is never transmitted to the server. \qedsymbol
\end{itemize}
\end{problem}

\section{Proposed Method}
\leavevmode{We propose \method (Collaborative Memory Bank), a memory-based method for accurate federated recommendation (Problem~\ref{pre:problem_def}).}
\leavevmode{To address existing limitations, we summarize the key technical challenges and our corresponding solutions as follows:}

\begin{itemize}[leftmargin=2em]
\item[\textbf{C1.}] \textbf{How can we summarize local interaction histories in a compact shared semantic space?} {We cluster pretrained item embeddings to define a shared coordinate system for collaboration and summarize local histories within it  (Section~\ref{sec:method:representation}).}

\item[\textbf{C2.}] \textbf{How can we organize clients into overlapping groups from these summaries?} 
{
We use LSH (Locality Sensitive Hashing) based on CWS (Consistent Weighted Sampling) to associate each client with multiple groups according to similarities among their cluster summaries (Section~\ref{sec:method:memory}).
}

\item[\textbf{C3.}] \textbf{How can we use overlapping client subsets to construct a client-specific collaborative memory?} 
We estimate {within-group cluster associations}, isolate subset-specific residuals from the global background, and aggregate memories from the client’s matched groups (Section~\ref{sec:method:memory}).

\item[\textbf{C4.}] \textbf{How can we use the retrieved memory to complement local recommendation?} {We derive collaborative scores from the retrieved memory, down-weight signals covered by local history, and combine them with local scores} (Section~\ref{sec:method:local}).
\end{itemize}

In the initialization phase (Algorithm~\ref{alg:initialization}), the server clusters the item representations once to construct a shared item-clustering structure, which each client $u$ downloads and stores locally. 
In the construction phase (Algorithm~\ref{alg:memory_bank}), each client $u$ summarizes its local interaction history $\Hist{u}$ as a compact sketch and uploads it to the server, which groups behaviorally similar clients using LSH and constructs the collaborative memory bank \(\mathcal{M}\). 
Finally, in the recommendation phase (Algorithm~\ref{alg:memory_retrieval}), the server uses each client’s signature to retrieve and aggregate relevant memories into a client-specific collaborative memory.
The client combines the resulting collaborative signals with its local scores to generate the final recommendation list \(L^{(u)}\) locally.

\subsection{Item Cluster-Based Client Sketching}
\label{sec:method:representation}

The existing methods encode collaborative knowledge in model states or item-level structures, tying knowledge sharing to repeated training or {\color{black}catalog-dependent communication.}
To address this, we represent such knowledge at the item-cluster level by projecting each local interaction history onto a shared item-cluster space and summarizing the resulting cluster-level behavior as a compact client sketch.
Because the sketch is defined over clusters rather than individual items, its size scales with the number of clusters, allowing the server to organize clients into behavior-conditioned collaborative groups without iterative client-server communication.

\smallsection{Semantic item clustering}
We construct a shared semantic coordinate system across clients by grouping items with similar content into semantic clusters. 
To this end, we use a precomputed $d$-dimensional language representation (LR) $\mathbf{e}_{i} \in \mathbb{R}^{d}$ for each item $i$, obtained by encoding its content (e.g., title) with a pretrained language model, {\color{black}and $L_2$-normalize the LRs to compare their semantic directions~\cite{Sheng0ZCWC25,JeonJGJ26}.}
Our initialization phase first performs clustering on item LRs only once at the server (Algorithm~\ref{alg:initialization}) as follows:
\begin{equation*}
    \mat{P} = \texttt{clustering}_{C}(\mat{E}), 
\end{equation*}
where $\mat{E} \in \mathbb{R}^{|\mathcal{I}|\times d}$ is the matrix of item LRs, $\mat{P} \in \mathbb{R}^{C \times d}$ is the matrix of centroids, and $C$ is the number of clusters. 
In this study, we used $K$-means clustering {\color{black}and $L_2$-normalized the resulting centroids for cosine-based item assignment}, but any embedding-based clustering algorithm can be used for this purpose. 
Then, the cluster assignment function for item $i$ is defined as follows:
\begin{equation}
\pi(i)
=
\argmax_{c\in\{1,\cdots,C\}}
\mathbf{e}_i^{\top}\mathbf{p}_c,
\label{eq:cluster_assignment}
\end{equation}
where the assignment $\pi(i)$ serves as a compact coordinate for item $i$ in the clustering space.
For each item–cluster pair $(i,c)$, we compute an affinity score $\mathbf{A}_{ic}$ as follows:
\begin{equation}
\mathbf{A}_{ic}
= \max\bigl(\cos(\mathbf{e}_{i},\mathbf{p}_{c}),0\bigr), \quad \mat{\tilde{A}} = \texttt{normalize}(\mat{A}). 
\label{eq:cluster_affinity}
\end{equation}

Note that we clip negative cosine similarities to zero to exclude unrelated clusters and focus the affinity scores solely on positively related clusters.
The item clustering result is represented as $\mathcal{S}\leftarrow (\mat{E},\mat{P},\pi,\mat{\tilde{A}})$, and each client downloads it from the server. 
Importantly, this communication, which linearly scales with the number of items, occurs only once at initialization.

\newcommand{\IndentedRequire}[1]{%
  \Statex \hspace{-0.6em}%
  \algorithmicrequire\ #1%
}
\newcommand{\IndentedEnsure}[1]{%
  \Statex \hspace{-0.6em}%
  \algorithmicensure\ #1%
}
\begin{algorithm}[t]
\caption{Initialization phase of \method}
\label{alg:initialization}
\small
\begin{algorithmic}[1]

\AlgPhase{\ding{182} Server}{It performs clustering on item LRs only once.}
\IndentedRequire{matrix $\mathbf{E}$ of item LRs, and number $C$ of clusters}

\State perform embedding-based clustering, i.e., $\mat{P} \gets \texttt{clustering}_{C}(\mat{E})$

\State compute cluster assignment function $\pi(\cdot)$ \Comment{{\scriptsize Eq.~\eqref{eq:cluster_assignment}}}

\State compute normalized item-cluster affinity matrix $\mat{\tilde{A}}$ \Comment{{\scriptsize Eq.~\eqref{eq:cluster_affinity}}}

\State store $\mathcal{S} \leftarrow (\mat{E},\mat{P},\pi,\mat{\tilde{A}})$ as item clustering structure

\AlgPhase{\ding{183} Client}{Each client obtains the clustering structure $\mathcal{S}$ from the server.}
\setcounter{ALG@line}{0} 
\State request \(\mathcal{S}\) from the server, then download and store it locally

\end{algorithmic}
\end{algorithm}

\smallsection{Client sketch generation}
Next, using the clustering result $\mathcal{S}$, we generate a compact sketch $\boldsymbol{\sigma}^{(u)}$ for each client $u$.
In the construction phase (Algorithm~\ref{alg:memory_bank}), the client first summarizes its interactions $\Hist{u}$ as a normalized cluster histogram:
\begin{equation}
\mathbf{\tilde{x}}_{c}^{(u)}
=
\frac{1}{|\Hist{u}|}
\sum_{i\in\Hist{u}}
\mathbb{I}(\pi(i)=c), 
\label{eq:client_histogram}
\end{equation}
where $\mathbb{I}(\cdot)$ returns 1 if the given predicate holds, and 0 otherwise.
It then computes the cluster-support vector $\mathbf{b}^{(u)}$ as follows:
\begin{equation}
\mathbf{b}_{c}^{(u)}
=
\mathbb{I}(\mathbf{\tilde{x}}_{c}^{(u)}>0).
\label{eq:binary_support}
\end{equation}

\begin{algorithm}[t]
\caption{Construction phase of \method}
\label{alg:memory_bank}
\small
\begin{algorithmic}[1]

\AlgPhase{\ding{182} Client}{Each client $u$ uploads its sketch $(\mathbf{h}_u,\mathbf{b}_u)$ to the server.}
\IndentedRequire{interaction history $\Hist{u}$ and CWS signature length $m$}

    \State summarize $\Hist{u}$ as normalized cluster histogram $\tilde{\mathbf{x}}^{(u)}$ using $\mathcal{S}$ \Comment{{\scriptsize Eq.~\eqref{eq:client_histogram}}}
    
    \State binarize $\tilde{\mathbf{x}}^{(u)}$ as cluster-support vector $\mathbf{b}^{(u)}$ \Comment{{\scriptsize Eq.~\eqref{eq:binary_support}}}  

    \State encode $\tilde{\mathbf{x}}^{(u)}$ into signature $\mathbf{h}^{(u)}$, i.e., $\mathbf{h}^{(u)} \gets \texttt{CWS}_{m}(\tilde{\mathbf{x}}^{(u)})$ \Comment{{\scriptsize Eq.~\eqref{eq:cws_signature}}}

    \State send $\boldsymbol{\sigma}^{(u)}\leftarrow(\mathbf{h}^{(u)}, \mathbf{b}^{(u)})$ to the server as the sketch for client $u$ 

\AlgPhase{\ding{183} Server}{Once all clients have uploaded their sketches, it constructs the 
  \hspace*{-\algorithmicindent}%
  \phantom{\textbf{\ding{183}\ }}%
  collaborative memory bank $\mathcal{M}$ using LSH.}
\IndentedRequire{$\{ \boldsymbol{\sigma}^{(u)} | u \in \mathcal{U} \}$ \& LSH parameters $b$ (bands) and $r$ (rows$/$band)}
\setcounter{ALG@line}{0}

\State construct LSH buckets $\mathcal{B}$, i.e., $\mathcal{B} \gets \texttt{LSH}_{b,r}(\{\mathbf{h}^{(u)} \mid \forall u\in\mathcal U\})$

\State discard singleton buckets in $\mathcal{B}$

\State compute global cluster conditional-support matrix $\mathbf{G}$ using $\mathbf{b}^{(u)}$ \Comment{{\scriptsize Eq.~\eqref{eq:global_association}}}
    
\For{\textbf{each} bucket $B\in\mathcal{B}$}
\State compute bucket cluster conditional-support matrices $\mathbf{Q}^{(B)}$ \Comment{{\scriptsize Eq.~\eqref{eq:bucket_association}}}

\State compute bucket collaborative memories $\mathbf{M}^{(B)}$ \Comment{{\scriptsize Eq.~\eqref{eq:bucket_memory}}}
\EndFor

\State store $\mathcal{M}\gets\{\mathbf{M}^{(B)} \mid B\in\mathcal{B}\}$ as collaborative memory bank

\end{algorithmic}
\end{algorithm}

Here, $\mathbf{\tilde{x}}_{c}^{(u)}$ measures the proportion of client $u$'s interactions associated with cluster $c$, whereas $\mathbf{b}_{c}^{(u)}$ records only whether such interactions are present.
Afterwards, we encode $\mathbf{\tilde{x}}^{(u)} \in \mathbb{R}^{C}$ into an $m$-dimensional signature vector $\mathbf{h}^{(u)} \in \mathbb{R}^{m}$, where $m \leq C$ and $\mathbf{h}^{(u)}$ is subsequently used to group similar clients at the server (Section~\ref{sec:method:memory}).
To this end, the signature $\mathbf{h}^{(u)}$ should faithfully represent $\tilde{\mathbf{x}}^{(u)}$ while ensuring that clients with similar vectors yield similar signatures.
As $\tilde{\mathbf{x}}^{(u)}$ is a nonnegative weighted vector, we employ CWS (Consistent Weighted Sampling)~\cite{Ioffe10} to construct $\mathbf{h}^{(u)}$, preserving the weighted Jaccard similarities $J_w$ between client vectors in expectation.
For clients $u$ and $v$, their  similarity $J_w$ is defined as:
\begin{equation*}\small
J_w(\mathbf{\tilde{x}}^{(u)},\mathbf{\tilde{x}}^{(v)})
=
\frac{
\sum_{c=1}^{C}
\min\!\left(\mathbf{\tilde{x}}_{c}^{(u)},\mathbf{\tilde{x}}_{c}^{(v)}\right)
}{
\sum_{c=1}^{C}
\max\!\left(\mathbf{\tilde{x}}_{c}^{(u)},\mathbf{\tilde{x}}_{c}^{(v)}\right)
}.
\label{eq:weighted_jaccard}
\end{equation*}

We use $m$ CWS hash functions $\{h_i\}$ with the same set of random seeds shared across all clients, as follows:
\begin{equation}
\mathbf{h}^{(u)} \leftarrow \texttt{CWS}_m(\mathbf{\tilde{x}}^{(u)})
\defeq
\left[
{\color{black}
h_1(\mathbf{\tilde{x}}^{(u)}),
\cdots,
h_m(\mathbf{\tilde{x}}^{(u)})
}
\right],
\label{eq:cws_signature}
\end{equation}
where this satisfies $P[h_i(\tilde{\mathbf{x}}^{(u)}) = h_i(\tilde{\mathbf{x}}^{(v)})] = J_w(\tilde{\mathbf{x}}^{(u)}, \tilde{\mathbf{x}}^{(v)})$, called collision property, 
meaning that the fraction of matching hash values across $m$ samples
provides an unbiased estimate of the  similarity (refer to more details in~\cite{Ioffe10}).
Here, \emph{``consistent''} refers to applying the same randomized sampling rule
across clients, such that a sample from their shared weighted portion is
selected identically under the same randomness.
Thus, clients with more similar $\tilde{\mathbf{x}}^{(u)}$ tend to produce more
similar signatures.
We define the client sketch as $\boldsymbol{\sigma}^{(u)}=(\mathbf{h}^{(u)},\mathbf{b}^{(u)})$, whose size is $O(m+C)$, significantly smaller than $O(|\mathcal{I}|)$ in general, and send it to the server.

\subsection{Collaborative Memory Bank Construction}
\label{sec:method:memory}

After receiving $\{\boldsymbol{\sigma}^{(u)}\}$ from all clients, we construct a collaborative memory bank $\mathcal{M}$ at the server (Algorithm~\ref{alg:memory_bank}).
To this end, we first group clients exhibiting similar behaviors based on the sketch.
However, clients may share different portions of their behavior with different clients. 
We therefore use an overlapping grouping structure to capture such partial similarities.

\begin{figure}[t]
    \centering
    \includegraphics[width=0.9\linewidth]{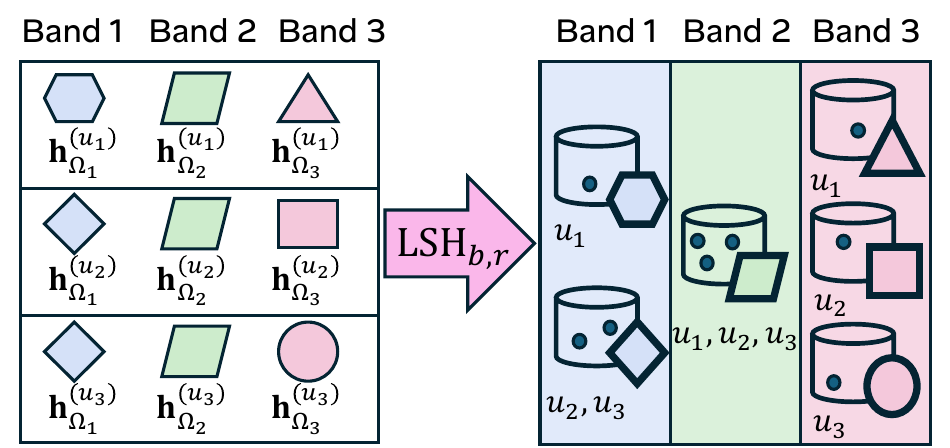}
    \caption{
    Example of LSH-based client indexing. Colors and shapes denote bands and bucket keys of sub-signatures, resp.
    Clients with identical keys within the same band share a bucket, resulting in overlapping client groups across bands.
    }
    \label{fig:LSH}
    \Description{Three clients have signatures divided into three bands. Equal shapes within a band map to the same bucket. The second band groups all three clients, while the first and third bands contain different overlapping subsets.}
\end{figure}

\smallsection{LSH-based client indexing}
To achieve this, we apply LSH (Locality Sensitive Hashing)~\cite{IndykM98, Broder97} to the CWS signatures as follows:
\begin{equation*}
    \mathcal{B} = \texttt{LSH}_{b,r}\bigl(\{\mathbf{h}^{(u)} \mid \forall u\in\mathcal U\}\bigr),
\end{equation*}
where $\mathcal{B}$ is the collection of band-specific buckets, $b$ is the number of bands, and $r$ is the number of signature entries (or rows) per band (i.e., $m=br$).
Specifically, it partitions each signature $\mathbf{h}^{(u)}\in\mathbb{R}^{m}$ as:
\begin{equation*}
    \mathbf{h}^{(u)} =  \left[\mathbf{h}^{(u)}_{\Omega_1}, \cdots, \mathbf{h}^{(u)}_{\Omega_b}\right],
\end{equation*}
where $\mathbf{h}^{(u)}_{\Omega_j} \in \mathbb{R}^{r}$ is its sub-signature for $j$-th band and $\Omega_j$ is the corresponding index range.
Each client $u$ is assigned to one bucket in each band $\Omega_j$, using $\mathbf{h}^{(u)}_{\Omega_j}$ as the key, and is thus indexed into $b$ buckets in total.
Treating each bucket as a group, a client may be grouped with different clients across bands, thereby producing overlapping groups (Figure~\ref{fig:LSH}). 
Moreover, since each bucket is determined by the sub-signature, it captures similarity under a partial view of client behavior.
\leavevmode{We discard singleton buckets (those with one client), as they provide no collaborative information.}

\smallsection{Collaborative memory construction}
We then construct a collaborative memory $\mathbf{M}^{(B)}$ for each LSH bucket $B$.
Each memory captures cluster associations that are particularly strong among the clients assigned to the bucket.
However, associations observed within a bucket may also include patterns common across all clients. 
To isolate bucket-specific knowledge, we compare the bucket-level associations with global associations estimated from all clients.

We first estimate the global association between clusters $i$ and $j$ as their co-occurrence across all clients:
\begin{equation}
\mat{G}_{ij}
=
\sum_{u\in\Users}
\vect{b}_{i}^{(u)}\cdot\vect{b}_{j}^{(u)},
\quad
\mat{\tilde{G}} = \texttt{normalize}(\mat{G}),
\label{eq:global_association}
\end{equation}
where $\mat{\tilde{G}} \in \mathbb{R}^{C \times C}$, and the diagonal entries of $\mat{G}$ are set to zero before normalization to exclude trivial self-co-occurrences.
\leavevmode{Similarly, we compute the association between clusters $i$ and $j$ within bucket $B$:}
\begin{equation}
\mat{Q}_{ij}^{(B)}
=
\sum_{u\in B}
\vect{b}_{i}^{(u)}\cdot\vect{b}_{j}^{(u)},
\quad
\mat{\tilde{Q}}^{(B)} = \texttt{normalize}(\mat{Q}^{(B)}).
\label{eq:bucket_association}
\end{equation}

To isolate bucket-specific knowledge, we remove the global associations from the bucket-specific associations as follows:
\begin{equation}
\mat{M}^{(B)}
=
\texttt{sparsify}_{R'}\bigl(\max(
\mat{\tilde{Q}}^{(B)}-\mat{\tilde{G}},
0
)\bigr). 
\label{eq:bucket_memory}
\end{equation}

We retain and sparsify only the positive residuals, as non-positive or weak residuals provide little bucket-specific support beyond the global associations.
We set $R'=2R$ to retain sufficient values before the additional sparsification used in Section~\ref{sec:memory:loca:ccm}.
For each bucket $B \in \mathcal{B}$, we compute $\mathbf{M}^{(B)}$ and store it into the bank $\mathcal{M}$.

\begin{algorithm}[t]
\caption{Recommendation phase of \method}
\label{alg:memory_retrieval}
\small
\begin{algorithmic}[1]

\AlgPhase{\ding{182} Client}{Each client $u$ requests collaborative memory from the server.}
\State {send} $\mathbf{h}^{(u)}$ to the server

\AlgPhase{\ding{183} Server}{It retrieves client-specific collaborative memory.}
\IndentedRequire{signature $\mathbf{h}^{(u)}$ and row-wise sparsity budget $R$}
\setcounter{ALG@line}{0}

\State look up matching LSH buckets $\mathcal{B}^{(u)}$, i.e., $\mathcal{B}^{(u)} \gets \texttt{LSHLookup}(\mathbf{h}^{(u)})$

\State retrieve bucket memories $\{\mathbf{M}^{(B)} \mid B\in\mathcal{B}^{(u)}\}$ from $\mathcal{M}$ 

\State compute client-specific collaborative memory $\mathbf{M}^{(u)}$ \Comment{{\scriptsize Eq.~\eqref{eq:client_memory}}}

\State {send} $\mathbf{M}^{(u)}$ to client $u$

\AlgPhase{\ding{184} Client}{Each client $u$ generates recommendations locally.}
\IndentedRequire{ interactions $\Hist{u}$, client memory $\mathbf{M}^{(u)}$, and fusion weight $\beta$}
\IndentedEnsure{recommendation list $L^{(u)}$}
\setcounter{ALG@line}{0}

\State compute cluster complement weights $\mathbf{w}^{(u)}$ \Comment{{\scriptsize Eq.~\eqref{eq:complement_weight}}}

\For{\textbf{each} candidate item $j \in \mathcal{C}^{(u)}$}\Comment{{\scriptsize$\mathcal{C}^{(u)}$ is a set of candidate items}}
    \State compute collaborative relation score $\boldsymbol{\rho}_j^{(u)}$  \Comment{{\scriptsize Eq.~\eqref{eq:relation_score}}}  
    \State compute local and final scores $\boldsymbol{\ell}_j^{(u)}$  and $\mathbf{s}^{(u)}_j$ \Comment{{\scriptsize Eqs.~\eqref{eq:local_score}--~\eqref{eq:final_score}}}
\EndFor

\State generate recommendation list $L^{(u)}$ based on $\mathbf{s}^{(u)}$ and \textbf{return} $L^{(u)}$

\end{algorithmic}
\end{algorithm}

\subsection{Recommendation with Client Memory}
\label{sec:method:local}

Once $\mathcal{M}$ is constructed, each client $u$ retrieves collaborative memory $\mathbf{M}^{(u)}$ relevant to its clustering-level behavior and uses it for local recommendation (Algorithm~\ref{alg:memory_retrieval}). 

\subsubsection{Client-specific collaborative memory retrieval}
\label{sec:memory:loca:ccm}
Each client first sends its signature $\mathbf{h}^{(u)}$ to the server to retrieve its $\mathbf{M}^{(u)}$.
Given $\mathbf{h}^{(u)}$, the server retrieves LSH buckets containing clients that are partially similar to client $u$, i.e., $\mathcal{B}^{(u)} = \texttt{LSHLookup}(\mathbf{h}^{(u)})$.
The server then retrieves the memory for each bucket $B \in \mathcal{B}^{(u)}$
from the memory bank $\mathcal{M}$, and computes $\mathbf{M}^{(u)}$ as follows:
\begin{equation}
\mat{M}^{(u)}
= \texttt{sparsify}_{R}\bigl(
\texttt{mean}(\{\mathbf{M}^{(B)} \mid B \in \mathcal{B}^{(u)}\})\bigr),
\label{eq:client_memory}
\end{equation}
where we further sparsify the averaged memory to make it more compact.
The resulting $\mathbf{M}^{(u)}$ thus provides compact collaborative knowledge tailored to the clustering-level behavior of client $u$ and is sent back to the client for local recommendation.

\subsubsection{Local recommendation on client}
After receiving $\mathbf{M}^{(u)}$ from the server, each client $u$ locally generates recommendations $L^{(u)}$ by combining local scores based on the semantics of consumed items in $\mathcal{H}^{(u)}$ with  relation scores derived from $\mathbf{M}^{(u)}$.

\smallsection{Collaborative relation score}
We first define the collaborative relation score $\boldsymbol{\rho}_j^{(u)}$ for candidate item $j$ by measuring how strongly it is supported by the cluster-level relations encoded in $\mathbf{M}^{(u)}$.
However, directly using $\mathbf{M}^{(u)}$ treats all cluster-level knowledge equally, even when some clusters are already well represented in $\Hist{u}$.
We therefore introduce cluster complement weights $\mathbf{w}^{(u)} \in \mathbb{R}^{C}$. 
Intuitively, less represented clusters in $\Hist{u}$ receive larger weights, while well-represented clusters receive smaller weights as follows:
\begin{equation}
\mathbf{w}_c^{(u)}
=
(\kappa_c^{(u)}-1)
/ (C-1),
\label{eq:complement_weight}
\end{equation}
where 
$\bar{\mathbf{e}}^{(u)} = \texttt{mean}(\{\mathbf{e}_i \mid i \in \Hist{u}\})$, and
$\kappa_c^{(u)}$ denotes the rank of $\cos(\mathbf{p}_c,\bar{\mathbf{e}}^{(u)})$ in the descending ordering of the $C$ clusters, with $1$ indicating the highest rank and $C$ the lowest.
In other words, higher-ranked clusters (i.e., those with $\kappa_c^{(u)}$ closer to $1$) receive smaller weights, whereas lower-ranked clusters receive larger weights.

Using $\mathbf{w}^{(u)}$, we next measure how strongly each candidate is supported by the client memory $\mathbf{M}^{(u)}$. 
Since both an interacted item and a candidate can be associated with multiple clusters, we compute their relation by combining all cluster-level relations according to their item-cluster affinities $\mathbf{A}$.
Let {\color{black}$\mathcal{C}^{(u)}$} denote the candidate set (e.g., all unseen items for user $u$). 
Specifically, for an interacted item $i \in \Hist{u}$ and a candidate item {\color{black}$j \in \mathcal{C}^{(u)}$}, we define the pairwise relation as:
\begin{equation}
\phi^{(u)}(i,j)
=
\sum_{a=1}^{C}\sum_{b=1}^{C}
{\color
{black}\mat{\tilde{A}}_{ia} \mat{M}_{ab}^{(u)} \mat{w}_{b}^{(u)} \mat{\tilde{A}}_{jb} =
\mat{\tilde{a}}_{i}^{\top}
\mat{M}^{(u)}
\mat{D}^{(u)}
\mat{\tilde{a}}_{j},
}
\label{eq:relation_contribution}
\end{equation}
where $\mathbf{D}^{(u)}=\texttt{diag}(\mathbf{w}^{(u)})$ is the diagonal matrix of cluster complement weights. 
Each term represents the support from item $i$ to candidate $j$ through cluster pair $(a,b)$, where $\mat{\tilde{A}}_{ia}$ and $\mat{\tilde{A}}_{jb}$ capture their cluster affinities, $\mat{M}_{ab}^{(u)}$ captures the cluster relation, and $\mat{w}_b^{(u)}$ emphasizes less represented clusters.

We assume that $\Hist{u}$ contains multiple unrelated interests, such that aggregating all interactions in $\Hist{u}$ for $\boldsymbol{\rho}_j^{(u)}$ may dilute the evidence relevant to candidate item \(j\). 
To preserve its strongest support, we use max aggregation as follows:
\begin{equation}
\boldsymbol{\rho}_j^{(u)}=
\max_{i\in\Hist{u}}
\phi^{(u)}(i,j).
\label{eq:relation_score}
\end{equation}

\smallsection{Fusion with local score}
We combine the relation score $\boldsymbol{\rho}_j^{(u)}$ with its local score $\boldsymbol{\ell}_j^{(u)}$ which is represented as follows:
\begin{equation}
\boldsymbol{\ell}_j^{(u)}
=
\cos\bigl(
\bar{\mathbf{e}}^{(u)},
\mathbf{e}_j
\bigr),
\label{eq:local_score}
\end{equation}
where $\bar{\mathbf{e}}^{(u)}$ is the averaged LRs of the items in $\Hist{u}$, and the score measures how well candidate item \(j\) aligns with the user’s local semantic preference, without collaborative knowledge.
The final score $\mathbf{s}^{(u)}_j$ is defined as follows:
\begin{equation}
\mathbf{s}^{(u)}_j
=
z^{(u)}\bigl(
\boldsymbol{\ell}^{(u)}_j
\bigr)
+
\beta \cdot
z^{(u)}\bigl(
\boldsymbol{\rho}_j^{(u)}
\bigr),
\label{eq:final_score}
\end{equation}
where $z^{(u)}(\cdot)$ is client-level standardization over all candidate items, placing the local and collaborative scores on comparable scales, and $\beta \geq 0$ controls the contribution of the collaborative relation score $\boldsymbol{\rho}_j^{(u)}$ relative to the local score $\boldsymbol{\ell}_j^{(u)}$. 
\leavevmode{Finally, we construct the recommendation list $L^{(u)}$ using the top-$K$ items ranked by $\mathbf{s}^{(u)}$.}

\subsection{Complexity Analysis}
\label{sec:method:complexity}

\begin{table}[t]
\centering
\caption{Time complexity of \method, where 
$|\mathcal{B}|$, $C$, $m$, $b$, $d$, and $R$ denote the numbers of buckets, clusters, hash functions, bands, dimensions, and sparsity budget respectively.
}
\label{tab:time_complexity}
\setlength{\tabcolsep}{6.5pt}
\renewcommand{\arraystretch}{1.2}
\small
\begin{tabular}{ccc}
\hline
\toprule
\multirow{2}{*}{\textbf{\method}} & \multicolumn{2}{c}{\bf Time complexity}                         \\
                            & Client                   & Server                           \\ \midrule
\bf Algorithm~\ref{alg:initialization}                 & -                        & $O(|\mathcal{I}|Cd)$                       \\
\bf Algorithm~\ref{alg:memory_bank}                 & $O(n_u + mC)$             & $O(C^2 (|\mathcal{U}| + |\mathcal{B}| \log{C}))$ \\
\bf Algorithm~\ref{alg:memory_retrieval}                 & $O(n_u(d+CR|\mathcal{C}^{(u)}|))$ & $O(bCR+C^2\log R)$             \\
\bottomrule
\hline
\end{tabular}
\end{table}

We analyze the time and communication complexity of \method. The results are summarized in Tables~\ref{tab:time_complexity} and~\ref{tab:communication_complexity}, with detailed proofs provided in the online appendix.
In this analysis, $|\mathcal{U}|$  and $|\mathcal{I}|$ are the numbers of users and items, resp. 
Let $n_u=|\mathcal{H}^{(u)}|$ and {\color{black}$|\mathcal{C}^{(u)}|$} be the number of candidate items for client $u$.

\smallsection{Time complexity}
Assuming all other parameters are fixed, Algorithm~\ref{alg:initialization} is dominated by item-cluster construction at the server, with runtime linear in $|\mathcal{I}|$.
In Algorithm~\ref{alg:memory_bank}, server-side LSH indexing scales linearly with $|\mathcal{U}|$ and $|\mathcal{B}|$, while client-side sketching takes $O(n_u+mC)$ time.
In Algorithm~\ref{alg:memory_retrieval}, server-side client-memory retrieval is independent of dataset size, while client-side scoring takes {\color{black}$O(n_u|\mathcal{C}^{(u)}|)$} time, typically with {\color{black}$n_u \ll |\mathcal{C}^{(u)}|$}.

\begin{table}[t]
\centering
\caption{Per-client communication complexities of \method and baselines. $\rho$ is the CoFedRec similar-group ratio, and $P_{\mathcal M}$ the number of non-item model parameters. Up and Down denote the uplink (client-to-server) and downlink (server-to-client) communication costs, respectively.}
\label{tab:communication_complexity}
\setlength{\tabcolsep}{3pt}
\renewcommand{\arraystretch}{1.2}
\resizebox{\linewidth}{!}{%
\small
\begin{tabular}{lcccccc}
\hline
\toprule
\textbf{Method}                      & \multicolumn{2}{c}{Up}                                                                      & \multicolumn{2}{c}{Down}                                                                  & \multicolumn{2}{c}{Rounds}                                                                         \\ \midrule
CoFedRec                             & \multicolumn{2}{c}{$O(|\mathcal{I}|d)$}                                                         & \multicolumn{2}{c}{$O(\rho |\mathcal{I}|d)$}                                                  & \multicolumn{2}{c}{$T$}                                                                            \\
PFedRec                              & \multicolumn{2}{c}{$O(|\mathcal{I}|d)$}                                                         & \multicolumn{2}{c}{$O(|\mathcal{I}|d)$}                                                       & \multicolumn{2}{c}{$T$}                                                                            \\
GPFedRec                             & \multicolumn{2}{c}{$O(|\mathcal{I}|d)$}                                                         & \multicolumn{2}{c}{$O(|\mathcal{I}|d)$}                                                       & \multicolumn{2}{c}{$T$}                                                                            \\
FedNCF                               & \multicolumn{2}{c}{$O(|\mathcal{I}|d + P_\mathcal{M})$}                                         & \multicolumn{2}{c}{$O(|\mathcal{I}|d + P_\mathcal{M})$}                                       & \multicolumn{2}{c}{$T$}                                                                            \\
FedRecGEL                            & \multicolumn{2}{c}{$O(|\mathcal{I}|d + P_\mathcal{M})$}                                         & \multicolumn{2}{c}{$O(|\mathcal{I}|d + P_\mathcal{M})$}                                       & \multicolumn{2}{c}{$T$}                                                                            \\
FedCIA                               & \multicolumn{2}{c}{$O(|\mathcal{I}|^2)$}                                                        & \multicolumn{2}{c}{$O(|\mathcal{I}|^2)$}                                                      & \multicolumn{2}{c}{$T$ or 1}                                                                       \\ \midrule
\multirow{3}{*}{\textbf{\method}} & \multicolumn{2}{c}{\bf Algorithm~\ref{alg:initialization}} & \multicolumn{2}{c}{\bf Algorithm~\ref{alg:memory_bank}} & \multicolumn{2}{c}{\bf Algorithm~\ref{alg:memory_retrieval}} \\\cline{2-7}
                                     & Up                                 & Down                                               & Up                                         & Down                                     & Up                                          & Down                                         \\
                                     & $-$                                    & $O(|\mathcal{I}|(d+C))$                                & $O(m+C)$                                       & $-$                                          & $O(m)$                                          & $O(CR)$                                          \\ 
                                     
                                     \bottomrule
                                     \hline
\end{tabular}
}
\end{table}

\newcommand{\cellp}[3]{#1{\scriptsize$\pm$#2 (#3)}}
\newcommand{\cellb}[3]{\textbf{#1}{\scriptsize$\pm$#2 (#3)}}
\newcommand{\cellu}[3]{\underline{#1}{\scriptsize$\pm$#2 (#3)}}

\newcommand{\cellbs}[3]{\textbf{#1}$^{*}${\scriptsize$\pm$#2 (#3)}}

\begin{table*}[t!]
\centering
\caption{
Performance in terms of HR@10 (or HR) and NDCG@10 (or ND) across six datasets. 
The best results are in \textbf{bold}, and the second-best results are \underline{underlined}. Numbers in parentheses indicate column-wise ranks.
\method achieves strong performance across all six datasets, attaining the best overall average rank among all methods.
}
\label{tab:exp:overall}
\setlength{\tabcolsep}{2pt}
\renewcommand{\arraystretch}{1.3}
\resizebox{\textwidth}{!}{%
\begin{tabular}{l|cc|cc|cc|cc|cc|cc|cc}
\hline
\toprule
&
\multicolumn{2}{c|}{\beauty}
&
\multicolumn{2}{c|}{\arts}
&
\multicolumn{2}{c|}{\movies}
&
\multicolumn{2}{c|}{\video}
&
\multicolumn{2}{c|}{\sports}
&
\multicolumn{2}{c|}{\industrial}
&
\multicolumn{2}{c}{Avg. Rank}
\\
 \textbf{Method}
& \textbf{HR} & \textbf{ND}
& \textbf{HR} & \textbf{ND}
& \textbf{HR} & \textbf{ND}
& \textbf{HR} & \textbf{ND}
& \textbf{HR} & \textbf{ND}
& \textbf{HR} & \textbf{ND}
& $\;$\textbf{HR} & \textbf{ND}
\\
\midrule
\texttt{Local}
& \cellu{.174}{.000}{2} & \cellb{.098}{.000}{1}
& \cellp{.384}{.000}{3} & \cellp{.209}{.000}{3}
& \cellp{.345}{.000}{4} & \cellp{.197}{.000}{4}
& \cellp{.393}{.000}{3} & \cellp{.225}{.000}{3}
& \cellu{.368}{.000}{2} & \cellp{.196}{.000}{3}
& \cellp{.362}{.000}{4} & \cellp{.206}{.000}{3}
& 3.0 & 2.8
\\
\midrule
FedNCF
& \cellp{.080}{.012}{8} & \cellp{.036}{.004}{7}
& \cellp{.230}{.018}{7} & \cellp{.123}{.010}{7}
& \cellp{.139}{.006}{7} & \cellp{.066}{.004}{7}
& \cellp{.286}{.018}{7} & \cellp{.161}{.012}{7}
& \cellp{.239}{.016}{6} & \cellp{.134}{.010}{6}
& \cellp{.365}{.007}{3} & \cellb{.216}{.003}{1}
& 6.3 & 5.8
\\
FedCIA
& \cellp{.086}{.006}{6} & \cellp{.046}{.003}{5}
& -- & --
& -- & --
& -- & --
& -- & --
& -- & --
& 6.0 & 5.0
\\
CoFedRec
& \cellp{.089}{.020}{5} & \cellp{.043}{.013}{6}
& \cellp{.266}{.005}{5} & \cellp{.141}{.002}{5}
& \cellp{.235}{.001}{6} & \cellp{.121}{.001}{6}
& \cellp{.330}{.004}{4} & \cellp{.185}{.002}{5}
& -- & --
& -- & --
& 5.0 & 5.5
\\
GPFedRec
& \cellp{.099}{.036}{4} & \cellp{.048}{.020}{3}
& \cellp{.263}{.004}{6} & \cellp{.140}{.003}{6}
& \cellp{.248}{.019}{5} & \cellp{.130}{.010}{5}
& \cellp{.329}{.003}{5} & \cellp{.185}{.003}{5}
& \cellp{.266}{.011}{5} & \cellp{.147}{.006}{5}
& -- & --
& 5.0 & 4.8
\\
PFedRec
& \cellp{.105}{.044}{3} & \cellp{.047}{.020}{4}
& \cellp{.327}{.005}{4} & \cellp{.183}{.003}{4}
& \cellp{.409}{.005}{3} & \cellp{.227}{.003}{3}
& \cellp{.328}{.005}{6} & \cellp{.190}{.002}{4}
& \cellp{.277}{.001}{4} & \cellp{.155}{.000}{4}
& -- & --
& 4.0 & 3.8
\\
FedRecGEL
& \cellp{.082}{.028}{7} & \cellp{.036}{.012}{7}
& \cellu{.387}{.004}{2} & \cellu{.219}{.003}{2}
& \cellb{.434}{.005}{1} & \cellb{.244}{.002}{1}
& \cellu{.400}{.005}{2} & \cellu{.229}{.003}{2}
& \cellp{.352}{.004}{3} & \cellu{.198}{.003}{2}
& \cellu{.373}{.002}{2} & \cellp{.204}{.004}{4}
& 2.8 & 3.0
\\

\midrule
\textbf{\method}
& \cellb{.178}{.011}{1} & \cellu{.093}{.005}{2}
& \cellbs{.414}{.002}{1} & \cellbs{.224}{.002}{1}
& \cellu{.417}{.004}{2} & \cellu{.235}{.002}{2}
& \cellbs{.456}{.003}{1} & \cellbs{.254}{.001}{1}
& \cellbs{.383}{.001}{1} & \cellbs{.204}{.001}{1}
& \cellbs{.379}{.001}{1} & \cellu{.213}{.001}{2}
& 1.2 & 1.5
\\
\bottomrule
\hline
\end{tabular}%
}
\vspace{0.3em}
\parbox{\textwidth}{
\footnotesize
\textbullet\; ``\textendash'' means that the corresponding baseline could not be evaluated on the dataset because its official implementation encountered an out-of-memory error under our setup. \\
\textbullet\; ``${*}$'' indicates a statistically significant improvement over the best baseline under a paired $t$-test across five random seeds ($p<0.05$). \\
\textbullet\; \texttt{Local} is deterministic given the fixed pretrained item representations and therefore has zero variance across runs.
}
\end{table*}

\smallsection{Communication complexity}
Table~\ref{tab:communication_complexity} summarizes the per-client communication complexity of each phase.
Algorithm~\ref{alg:initialization} requires a one-time catalog-dependent download of the shared item-side information.
{\color{black}
After the memory bank is constructed, each client uploads its compact signature $\vect{h}^{(u)}$ and receives a sparse client-specific memory $\mat{M}^{(u)}$ during recommendation.
These messages incur $O(m)$ uplink and $O(CR)$ downlink payloads, respectively, both substantially smaller than $O(|\Items|)$ by operating in the compact cluster space rather than the full item catalog.}
We provide a detailed comparison with those methods in Section~\ref{sec:exp:communication}.

\section{Experiments}
\leavevmode{We conducted experiments to answer the following questions:}

\begin{itemize}[leftmargin=2em]
    \item[\textbf{Q1.}] \textbf{Recommendation performance (Section~\ref{sec:exp:performance}).}
    {How accurate is \method compared with federated baselines?}
    
    \item[\textbf{Q2.}] \textbf{Collaborative information construction (Section~\ref{sec:exp:collaborative}).}
    Does \method construct meaningful collaborative information by representing and grouping clients in the shared semantic space?

    \item[\textbf{Q3.}] \textbf{Complementarity of client memory (Section~\ref{sec:exp:mechanism}).}
    Does the client collaborative memory provide complementary recommendation signals beyond local recommendation?
    
    \item[\textbf{Q4.}] \textbf{Adaptation to evolving client behavior (Section~\ref{sec:exp:adaption}).}
    How effectively can the client-specific collaborative memory be adapted to evolving client behavior without reconstructing the collaborative memory bank?

    \item[\textbf{Q5.}] \textbf{Communication efficiency (Section~\ref{sec:exp:communication}).}
    How much communication does \method require to construct and retrieve collaborative memories compared with existing federated recommendation methods?
    
\end{itemize}

\begin{table}[t!]
\centering
\small
\caption{Statistics of real-world recommendation datasets.}
\label{tab:datasets}
\setlength{\tabcolsep}{9pt}
\begin{tabular}{lrrrr}
\hline
\toprule
\textbf{Dataset} & \textbf{\# Users} & \textbf{\# Items} & \textbf{\# Inter.} & \textbf{Spars.} \\
\midrule
\beauty       & 253 & 356  & 2,535 & 97.19\% \\
\arts       & 11,734 & 8,319  & 188,984 & 99.81\% \\
\movies     & 10,409 & 10,113 & 430,603 & 99.59\% \\
\video      & 11,658 & 5,335  & 196,508 & 99.68\% \\
\sports     & 17,906 & 11,247 & 267,830 & 99.87\% \\
\industrial & 50,985 & 25,847 & 412,938 & 99.97\% \\
\bottomrule
\hline
\end{tabular}
\end{table}
Note that we report results for all datasets in Table~\ref{tab:exp:overall} but focus on \texttt{Arts}, \texttt{Movies}, and \texttt{Video} in subsequent analyses due to space constraints; results for the remaining datasets show similar trends and are provided in the online appendix.

\subsection{Experimental Settings}
\label{sec:experimental_setup}

\smallsection{Datasets}
We used six real-world datasets~\cite{HouLFHYCM26}, whose statistics are summarized in Table~\ref{tab:datasets}. 
These datasets cover diverse product categories in Amazon Reviews\footnote{\url{https://amazon-reviews-2023.github.io}}, namely \beauty, \arts, \movies, \video, \sports, and \industrial.

\smallsection{Baselines} We compared \method with five model-based methods, including FedNCF~\cite{PerifanisE22}, PFedRec~\cite{ZhangL0YZZY23}, GPFedRec~\cite{ZhangL0ZYY24}, CoFedRec~\cite{HeLKH24}, and FedRecGEL~\cite{YuFLZWC26}, which cover federated neural collaborative filtering, personalized collaboration under client heterogeneity, and improved generalization of shared item embeddings. We also included FedCIA~\cite{Han0XLGZGL25}, a relation-based method that aggregates collaborative information through item-level relations while preserving client-specific representation spaces.

\smallsection{Training and evaluation protocol}
Following recent studies~\cite{ZhangL0YZZY23, HeLKH24}, we adopted a timestamp-based leave-last-out splitting, where each user's most recent interaction was used for testing, the second most recent interaction for validation, and all remaining interactions for training.
For each user, we constructed a candidate set consisting of one ground-truth item and 99 randomly sampled negative items and ranked these items for evaluation~\cite{ZhangL0YZZY23, HeLKH24}.
We used Hit Rate (HR@10) and Normalized Discounted Cumulative Gain (NDCG@10) as evaluation metrics.
We selected hyperparameters based on HR@10 on the validation set and reported the test results averaged over five runs with different random seeds.

{\color{black}
\smallsection{Language representations}
{We precomputed item embeddings for \method by encoding each item's title with the pretrained language model \texttt{text-embedding-3-large}~\cite{abs-2201-10005,Sheng0ZCWC25}.}

We provide details on datasets, hyperparameter tuning, and other experimental settings in the online appendix.
}

\subsection{Recommendation Performance}
\label{sec:exp:performance}

We compare the overall recommendation performance of \method with the baselines in Table~\ref{tab:exp:overall}. 
Across datasets, our method achieves strong HR@10 and NDCG@10 performance, ranking first in 8 of the 12 dataset-metric pairs and second in the remaining four. 
It therefore achieves the best average ranks of 1.2 for HR@10 and 1.5 for NDCG@10.
{\color{black}
Importantly, \method also outperforms \texttt{Local}, which uses the same item LRs and local scoring function without collaborative information,
in 11 of 12 cases.
}
This shows that the gains come from the retrieved collaborative memory beyond the item LRs.
Note that some baselines could not be evaluated on the larger datasets due to memory-intensive exchange structures or repeated communication.
\leavevmode{Overall, these results show that the collaborative memory bank stores and retrieves useful collaborative knowledge without iterative client-server communication.}

\subsection{Collaborative Information Construction}
\label{sec:exp:collaborative}
We examine how CollabBank constructs meaningful collaborative information. We first test whether semantic item clusters provide a shared space for client behaviors (Figure~\ref{fig:semantic_coordinate_permutation}). We then test whether CWS-based LSH preserves behavioral similarity during client grouping (Figure~\ref{fig:exp:jaccard_collision}). 
Finally, we evaluate the recommendation utility of the resulting memories (Figure~\ref{fig:exp:prior_ablation}). 

\subsubsection{Semantic item-cluster assignments}
\begin{figure}[t]
    \centering
    \includegraphics[width=0.82\linewidth,trim=0 9bp 0 9bp,clip]{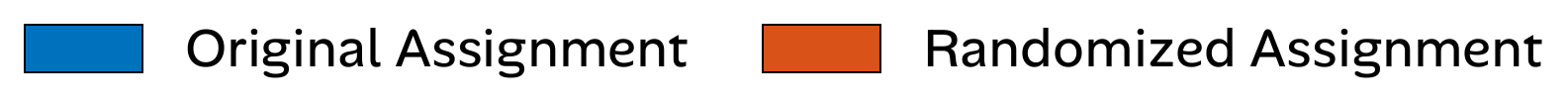}
    \\[2pt]
    \begin{subfigure}[c]{0.028\linewidth}
        \raisebox{4.5mm}        {\includegraphics[width=\linewidth]{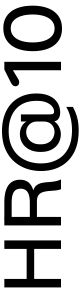}}
    \end{subfigure}
    \begin{subfigure}[c]{0.31\linewidth}
        \includegraphics[width=\linewidth]{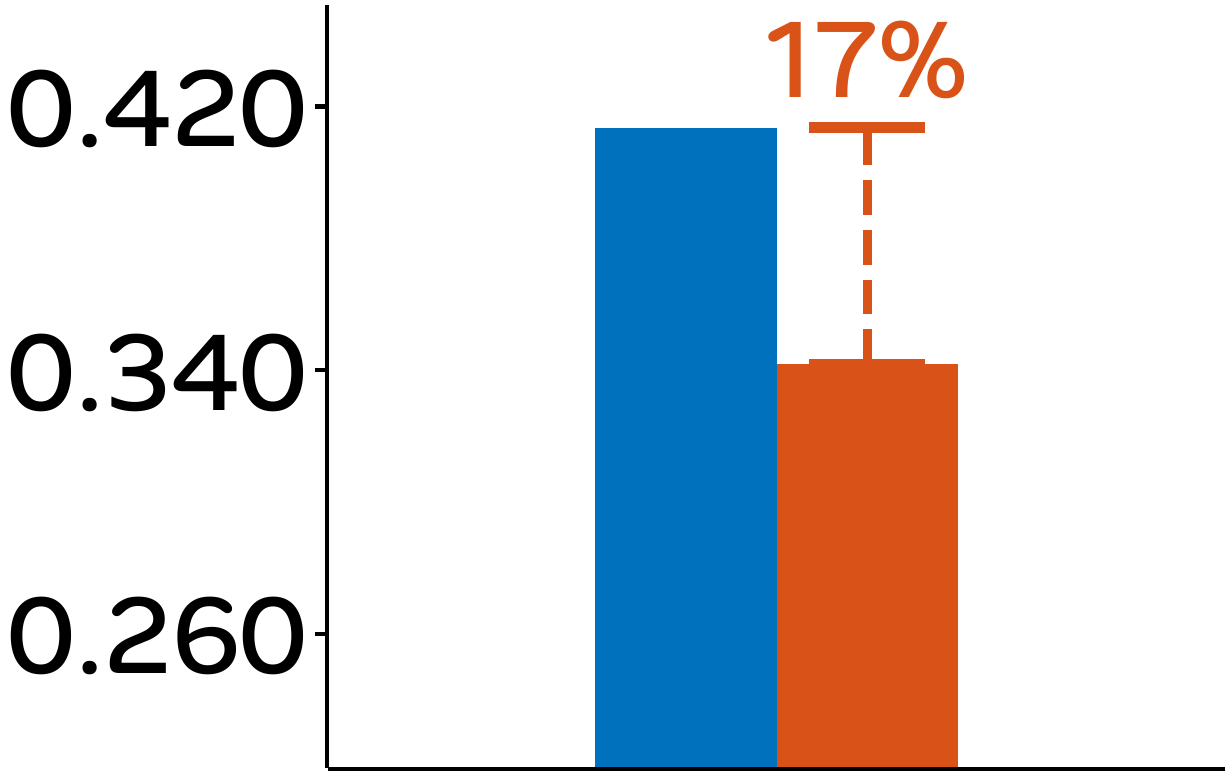}
        \captionsetup{margin={0mm,4mm}}
        \caption{Arts}
    \end{subfigure}
    \begin{subfigure}[c]{0.31\linewidth}
        \includegraphics[width=\linewidth]{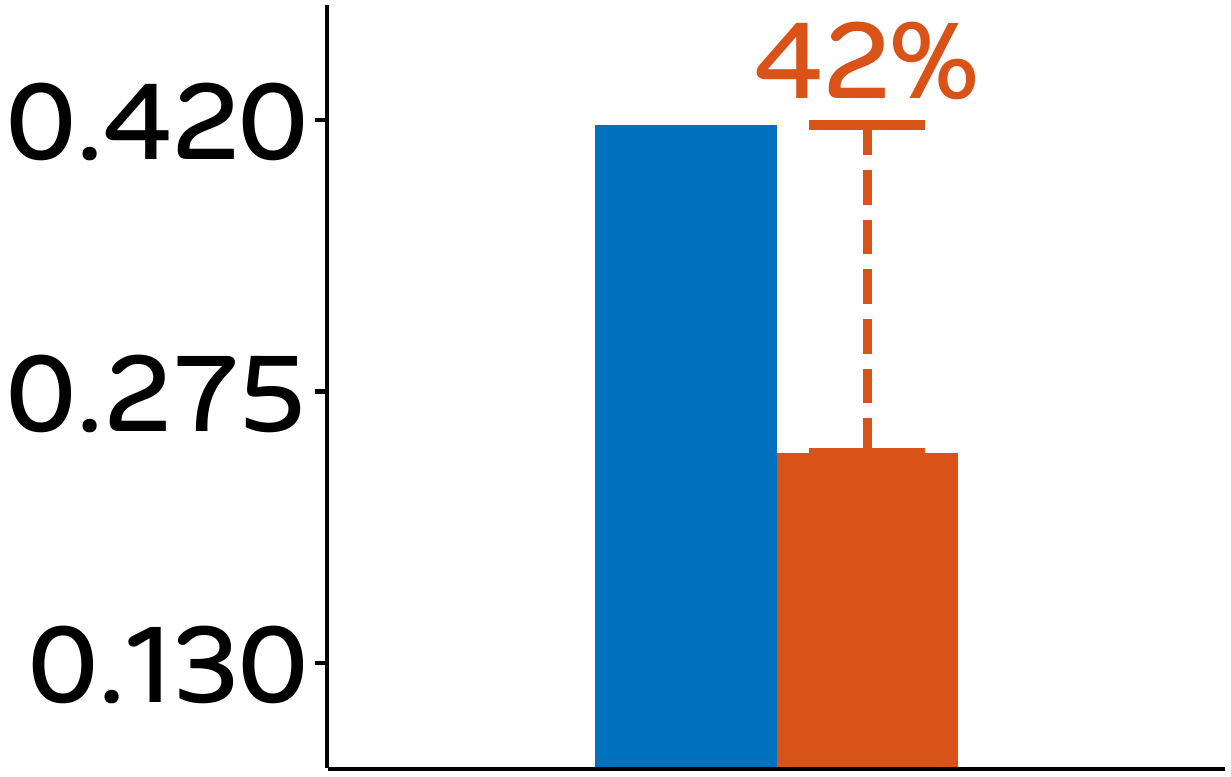}
        \captionsetup{margin={0mm,4mm}}
        \caption{Movies}
    \end{subfigure}
    \begin{subfigure}[c]{0.31\linewidth}
        \includegraphics[width=\linewidth]{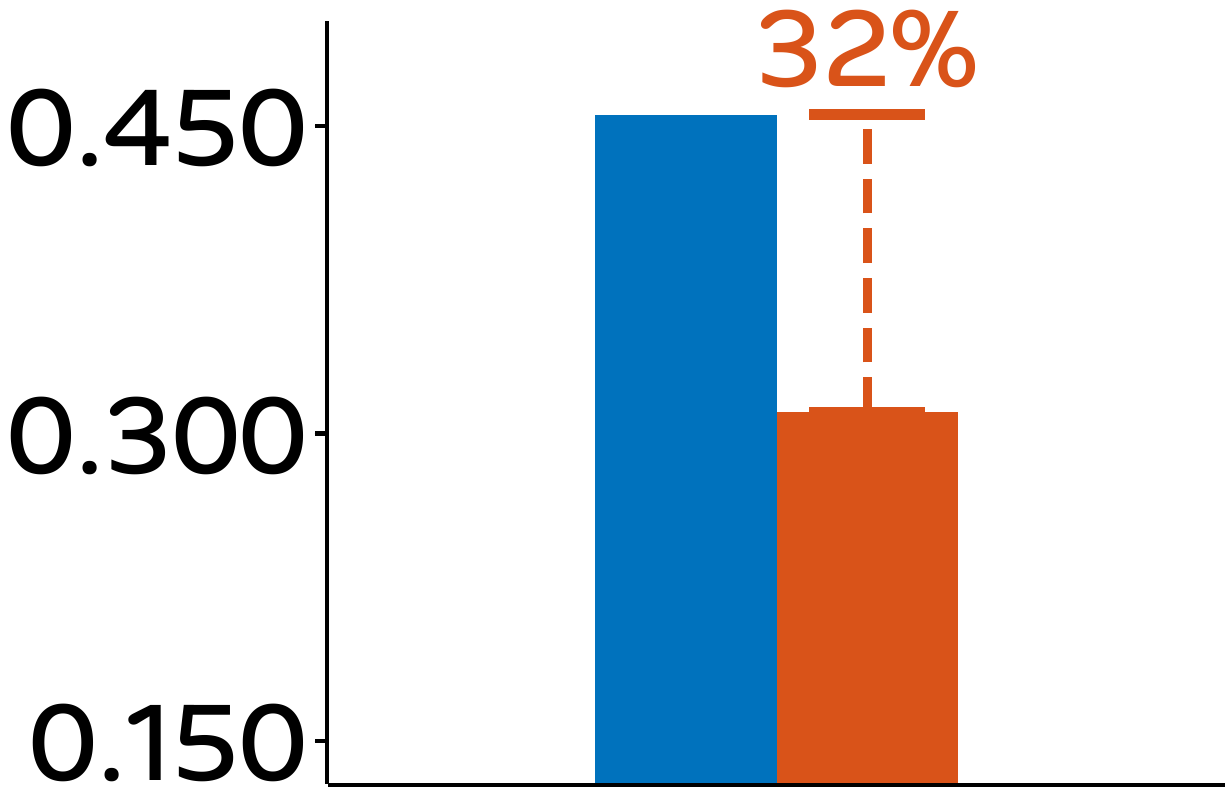}
        \captionsetup{margin={0mm,4mm}}
        \caption{Video}
    \end{subfigure}
    \caption{
    Recommendation performance under original and randomized item-cluster assignments. Randomization consistently degrades performance, highlighting the role of semantic item-cluster assignments in providing a common coordinate for collaborative information construction.
    }    \label{fig:semantic_coordinate_permutation}
    \Description{Three paired bar charts compare HR@10 for original and randomized assignments on Arts, Movies, and Video. The reductions from randomization are 17, 42, and 32 percent, respectively.}
\end{figure}

We first examine whether semantic item-cluster assignments provide a meaningful shared basis for constructing collaborative information across clients, rather than merely compressing the item space. 
{\color{black}
To isolate the effect of semantic structure, we randomize the item-cluster
assignments while preserving the number of items assigned to each cluster
and keeping the subsequent collaborative memory construction pipeline unchanged.
}
As shown in Figure~\ref{fig:semantic_coordinate_permutation}, randomized assignments consistently degrade recommendation performance across the evaluated datasets. This result indicates that preserving semantic item-cluster structure is important for collaborative information construction.

\subsubsection{Behavior-preserving LSH indexing} 
\begin{figure}[t]
    \centering
    \includegraphics[width=0.48\linewidth,trim=0 9bp 0 4bp,clip]{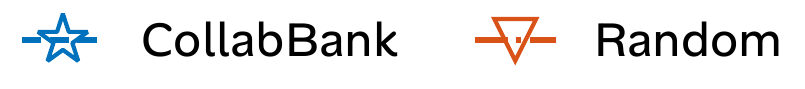}
    \\[-2mm]
    \begin{subfigure}[t]{0.05\linewidth}    
        {\includegraphics[width=\linewidth]{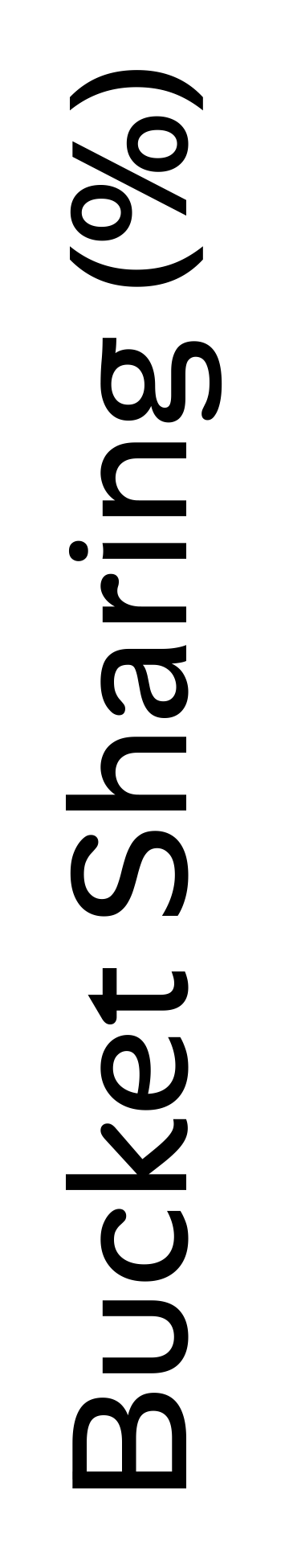}}
    \end{subfigure}
    \begin{subfigure}[t]{0.303\linewidth}
        \includegraphics[width=\linewidth]{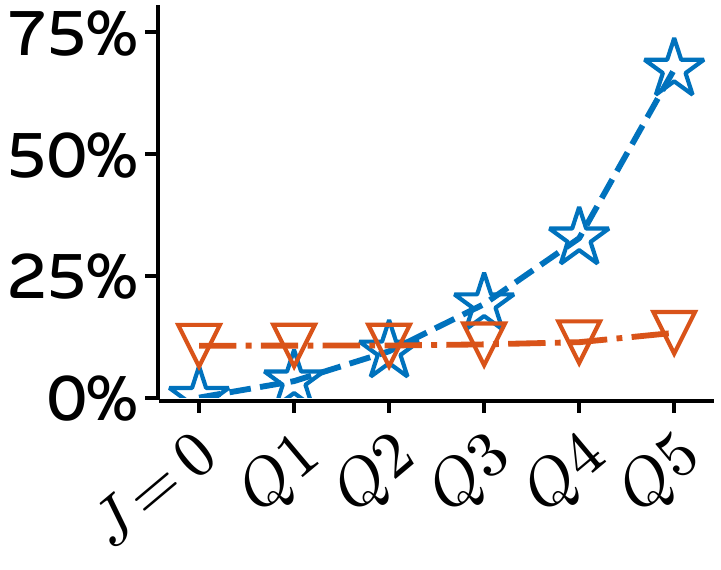}
        \captionsetup{margin={4mm,0mm}}
        \caption{Arts}
    \end{subfigure}
    \begin{subfigure}[t]{0.303\linewidth}
        \includegraphics[width=\linewidth]{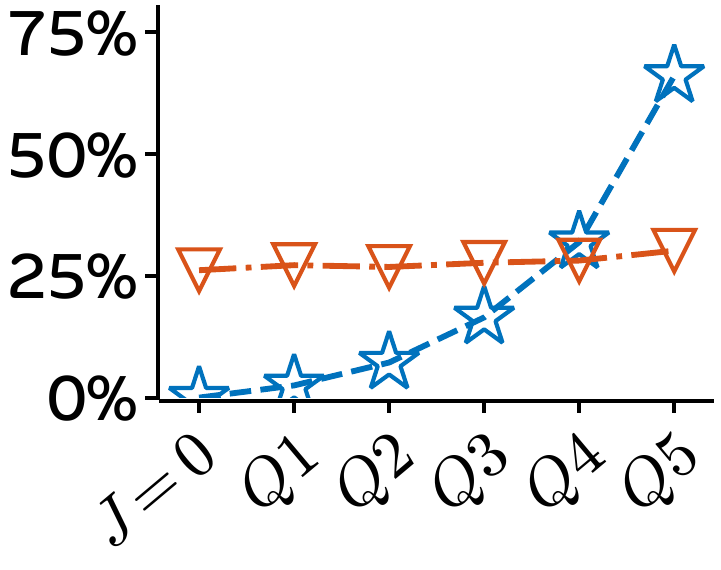}
        \captionsetup{margin={4mm,0mm}}
        \caption{Movies}
    \end{subfigure}
    \begin{subfigure}[t]{0.303\linewidth}
        \includegraphics[width=\linewidth]{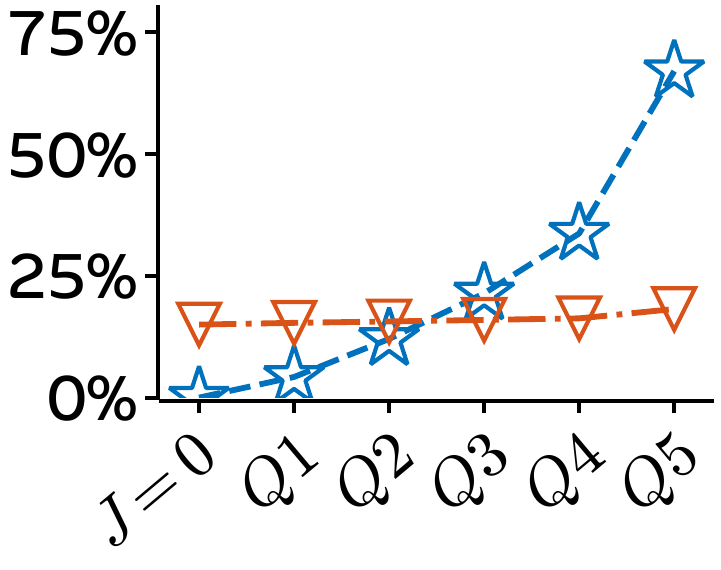}
        \captionsetup{margin={4mm,0mm}}
        \caption{Video}
    \end{subfigure}
    \caption{
    Bucket sharing across behavioral similarity levels under LSH and random grouping. As behavioral similarity increases, clients are more likely to share an LSH bucket, unlike random grouping, indicating that LSH groups clients by behavioral similarity.
    }
    \label{fig:exp:jaccard_collision}
    \Description{Three line charts plot bucket sharing percentage against zero weighted Jaccard similarity and five increasing similarity quantiles. The CollabBank curves rise toward the highest quantile, whereas the random grouping curves remain relatively flat.}
\end{figure}

{
We next examine whether LSH groups behaviorally similar clients together under the semantic item-cluster assignments established above.
Since CWS signatures are generated from the resulting client cluster representations, we first verify whether their behavioral similarity is preserved in the signatures.
We use weighted Jaccard similarity to measure behavioral similarity between client cluster representations and the fraction of matching entries to measure CWS signature similarity.
The two measures are strongly aligned, with an average Spearman correlation of 0.98 across datasets.
We then examine whether this preserved similarity translates into LSH bucket sharing.}
For LSH analysis, we separate client pairs with zero similarity (\(J=0\)) and divide the remaining pairs into five quantiles (\(Q1\)–\(Q5\)), ordered from low to high similarity. 
We then measure the fraction of pairs in each group that share at least one bucket. 
As shown in Figure~\ref{fig:exp:jaccard_collision}, bucket sharing increases from \(Q1\) to \(Q5\), while random grouping shows no such pattern. 
These results show that CWS-based LSH preserves behavioral similarity for client grouping. {We further provide LSH bucket statistics in the online appendix.}

\subsubsection{Client-specific collaborative memory}
\begin{figure}[t]
    \centering
    \includegraphics[width=0.68\linewidth,trim=0 16bp 0 13bp,clip]{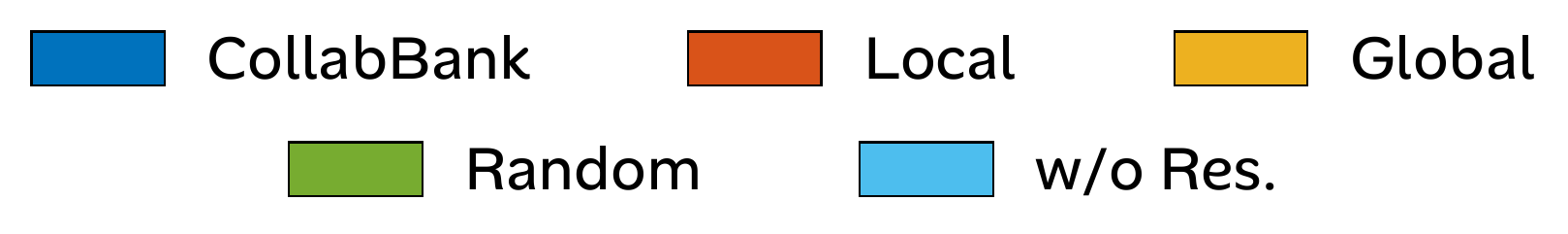}
    \\[2pt]
    \begin{subfigure}[c]{0.0275\linewidth}
        \raisebox{4mm}{\includegraphics[width=\linewidth]{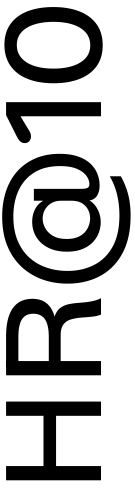}}
    \end{subfigure}
    \begin{subfigure}[c]{0.31\linewidth}
        \includegraphics[width=\linewidth]{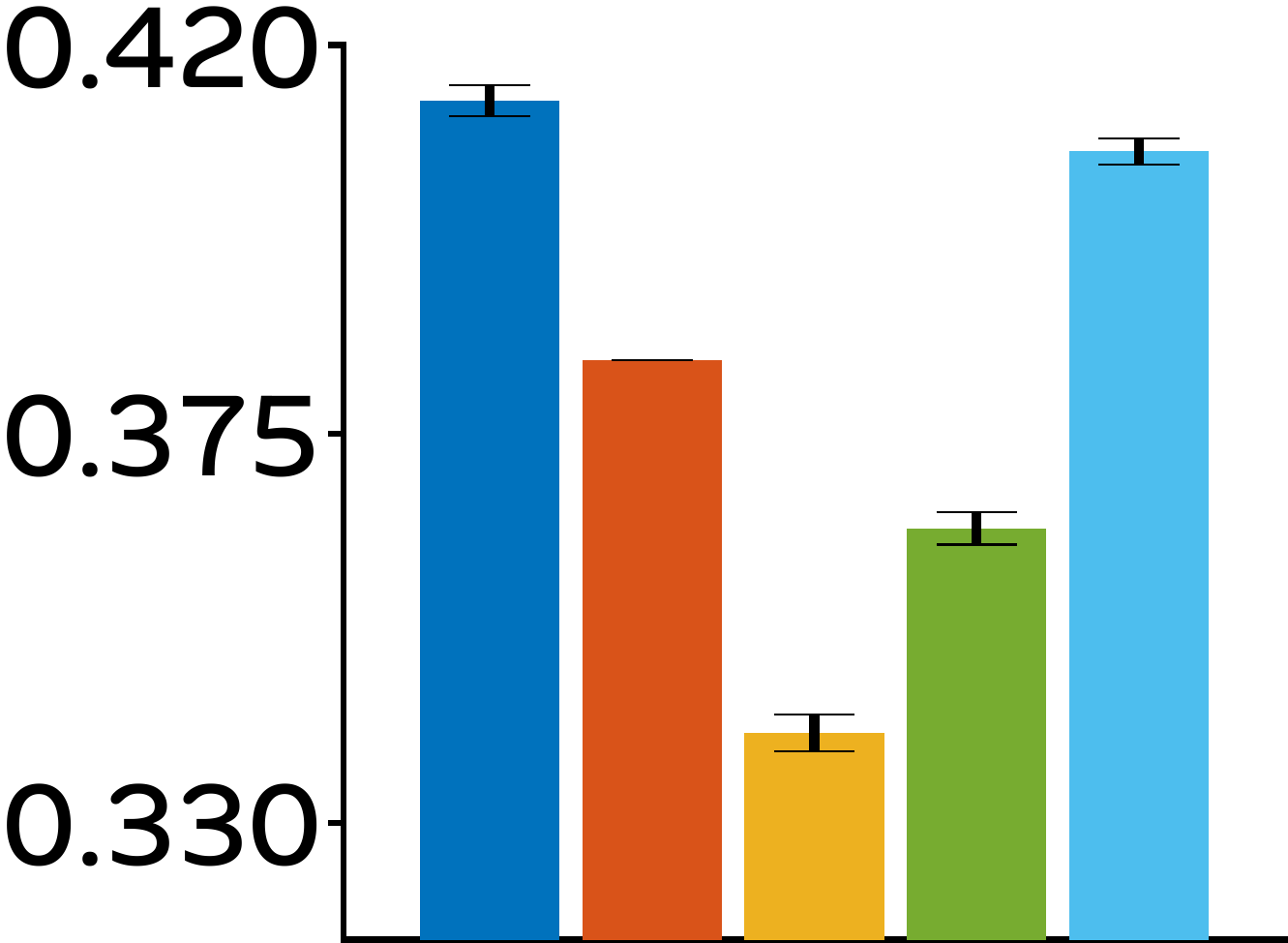}
        \captionsetup{margin={0mm,4mm}}
        \caption{Arts}
    \end{subfigure}
    \begin{subfigure}[c]{0.31\linewidth}
        \includegraphics[width=\linewidth]{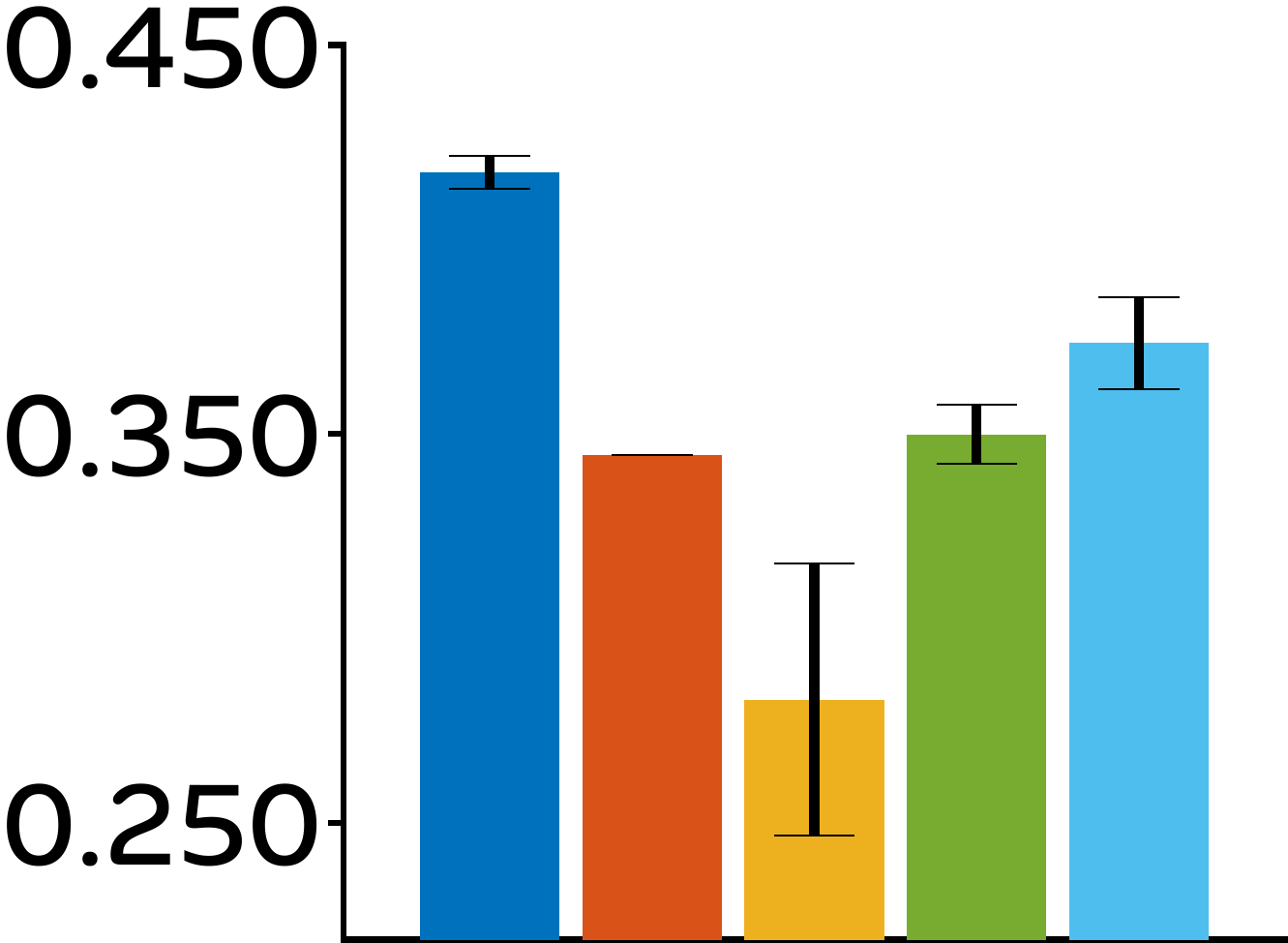}
        \captionsetup{margin={0mm,4mm}}
        \caption{Movies}
    \end{subfigure}
    \begin{subfigure}[c]{0.31\linewidth}
        \includegraphics[width=\linewidth]{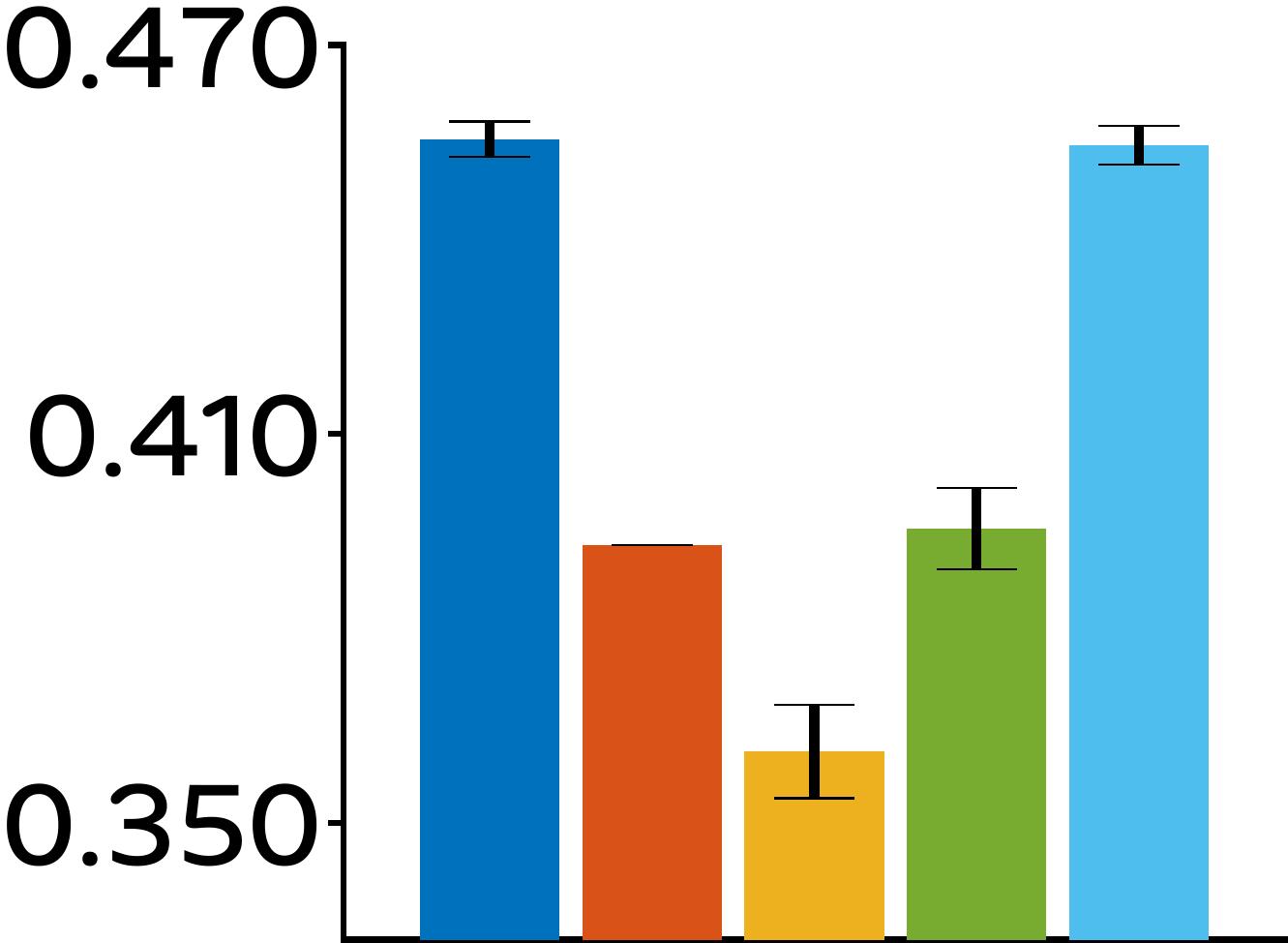}
        \captionsetup{margin={0mm,4mm}}
        \caption{Video}
    \end{subfigure}
    \caption{
    Recommendation performance across collaborative memory variants. Aggregating memories from LSH buckets that group behaviorally similar clients improves recommendation, with further gains from residualization.
    }
    \label{fig:exp:prior_ablation}
    \Description{Three bar charts compare HR@10 for CollabBank, Local, Global, Random, and the variant without residualization on Arts, Movies, and Video. CollabBank has the tallest bar in each chart. Error bars show variation across runs.}
\end{figure}

We examine whether our collaborative memory design improves recommendation performance. 
We compared \method with four variants. 
\texttt{Local} uses only the local semantic score in Eq.~\eqref{eq:local_score}, \texttt{Global} uses {\color{black}associations estimated} from all clients, and 
{\color{black}\texttt{Random} constructs memories from randomly assigned buckets while preserving the original bucket-size statistics.}
\texttt{w/o Res.} uses the same LSH buckets but does not remove global associations. 
As shown in Figure~\ref{fig:exp:prior_ablation}, \method and \texttt{w/o Res}. consistently outperform \texttt{Local}, \texttt{Global}, and \texttt{Random}, showing the importance of behavior-conditioned grouping. 
\texttt{Global} performs worse than \texttt{Local}, while \texttt{Random} 
{shows inconsistent performance relative to \texttt{Local},} indicating that collaboration without behavior conditioning can introduce irrelevant information. 
\method further improves over \texttt{w/o  Res.}, showing that residualization helps isolate bucket-specific collaborative information from global patterns.

\subsection{Complementarity of Client Memories}
\label{sec:exp:mechanism}
We analyze how well our client-specific collaborative memory complements local recommendation. 
\subsubsection{Decomposition of recommendation gains}

\begin{figure}[t]
    \centering
    \includegraphics[width=0.85\linewidth,trim=0 0bp 0 24bp,clip]{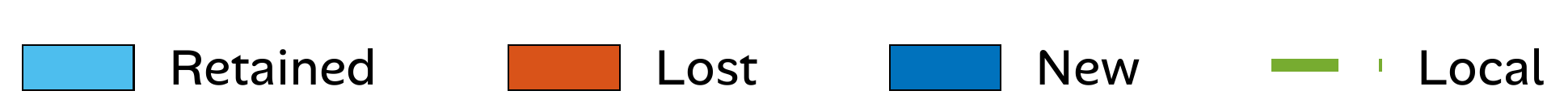}
    \\[4pt]
    \begin{subfigure}[c]{0.0275\linewidth}
        \raisebox{7mm}{\includegraphics[width=\linewidth]{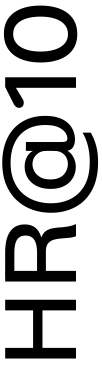}}
    \end{subfigure}
    \begin{subfigure}[c]{0.313\linewidth}
        \includegraphics[width=\linewidth]{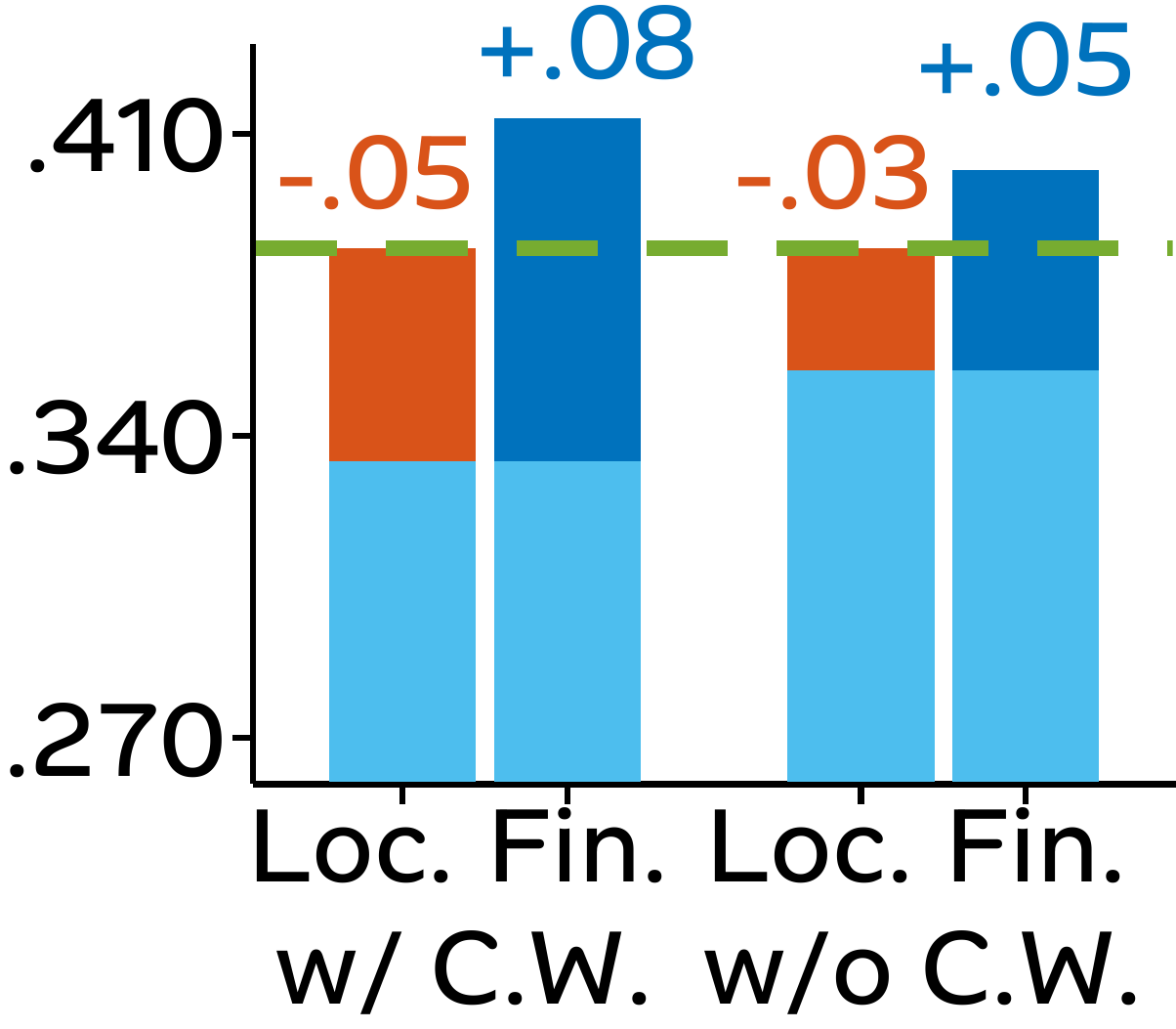}
        \captionsetup{margin={0mm,4mm}}
        \caption{Arts}
    \end{subfigure}
    \begin{subfigure}[c]{0.313\linewidth}
        \includegraphics[width=\linewidth]{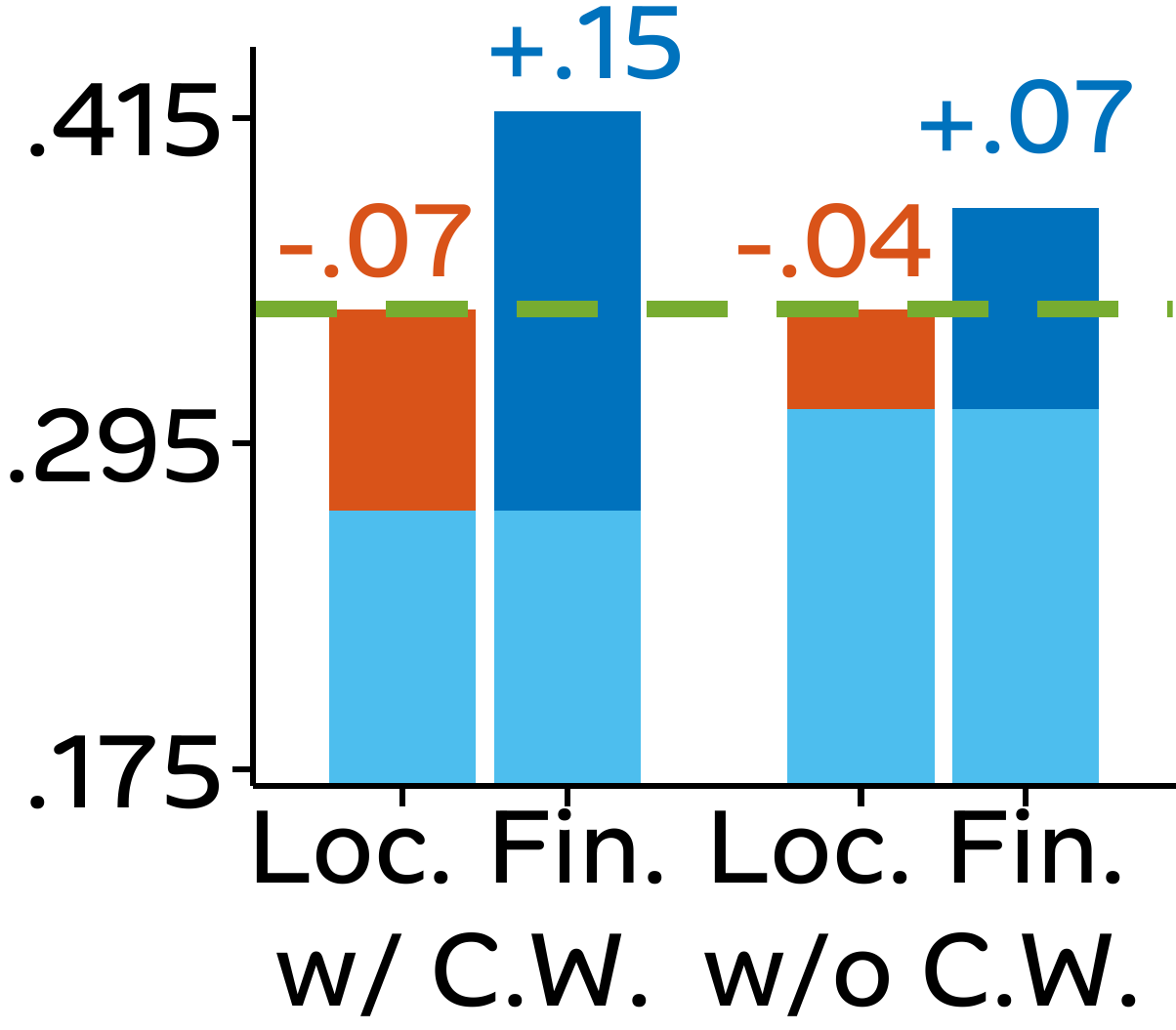}
        \captionsetup{margin={0mm,4mm}}
        \caption{Movies}
    \end{subfigure}
    \begin{subfigure}[c]{0.313\linewidth}
        \includegraphics[width=\linewidth]{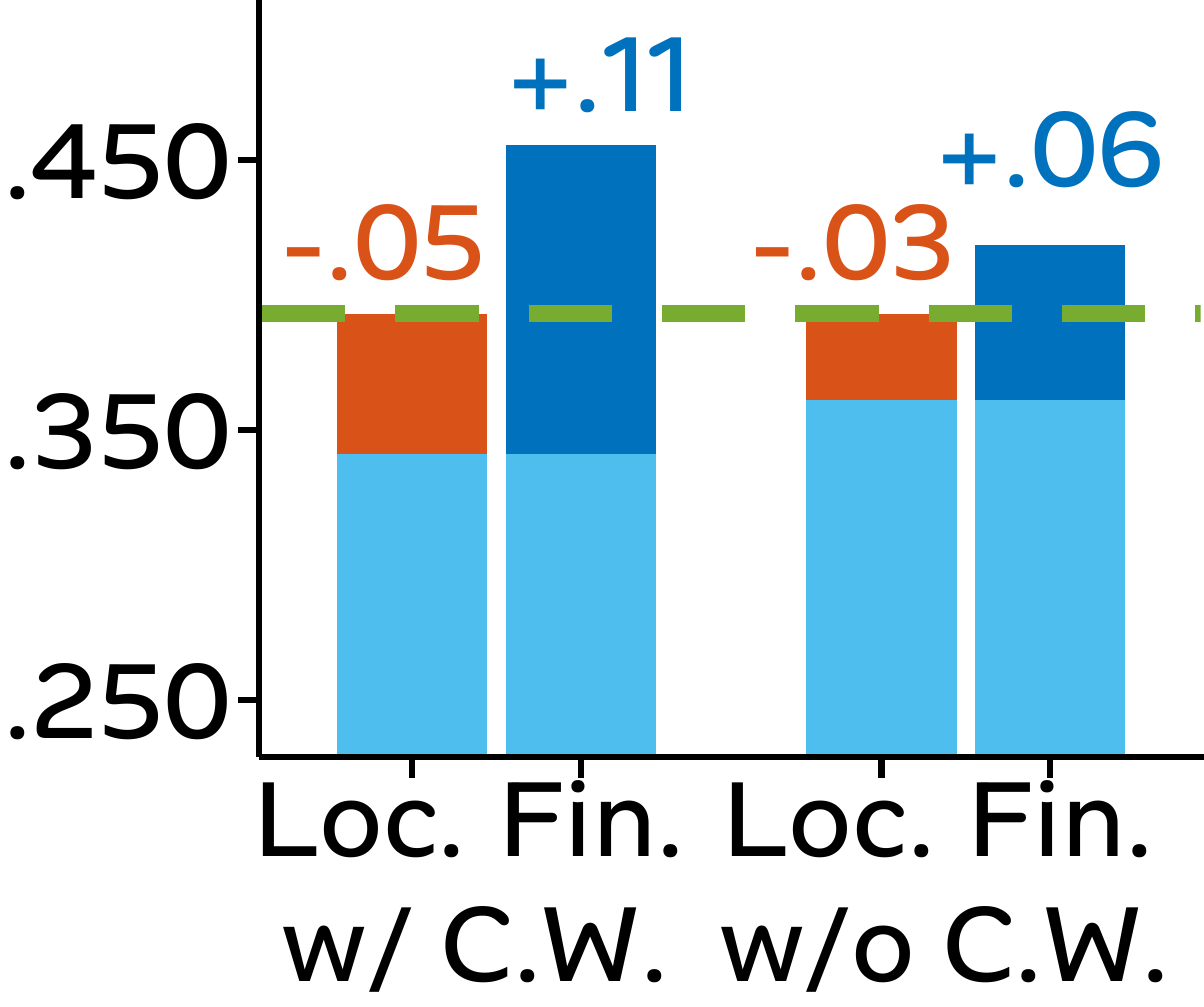}
        \captionsetup{margin={0mm,4mm}}
        \caption{Video}
    \end{subfigure}
    \caption{
    Decomposition of recommendation gains with and without complement weights. Collaborative relation scoring yields more new than lost hits, while complement weights further increase the net gain.
    }
    \label{fig:exp:hit_decomposition}
    \Description{For each dataset, paired stacked bars compare local and fused hits with and without complement weights. Retained, lost, and new hits are distinguished by color. The dashed line indicates local performance, and numeric labels indicate the lost and new hits.}
\end{figure}
{\color{black}
Figure~\ref{fig:exp:hit_decomposition} decomposes hits after fusion into
retained, lost, and new hits, corresponding to locally correct hits that are preserved, displaced, or newly introduced by collaborative memory, respectively.
The net gain is the difference between new and lost hits.
}
Across all datasets, new hits exceed lost hits, showing that collaborative relation scoring recovers relevant items missed by the local model.
Cluster complement weighting (C.W.) further increases this gap, indicating that it suppresses collaborative
information redundant with local history and promotes information that better
complements local recommendations.

\subsubsection{Sensitivity to the fusion weight}
\begin{figure}[t]
    \begin{subfigure}[c]{0.0275\linewidth}
        \raisebox{9.5mm}{\includegraphics[width=\linewidth]{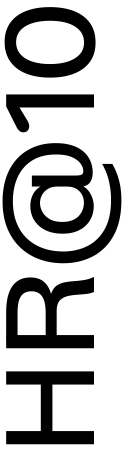}}
    \end{subfigure}
    \begin{subfigure}[c]{0.31\linewidth}
        \includegraphics[width=\linewidth]{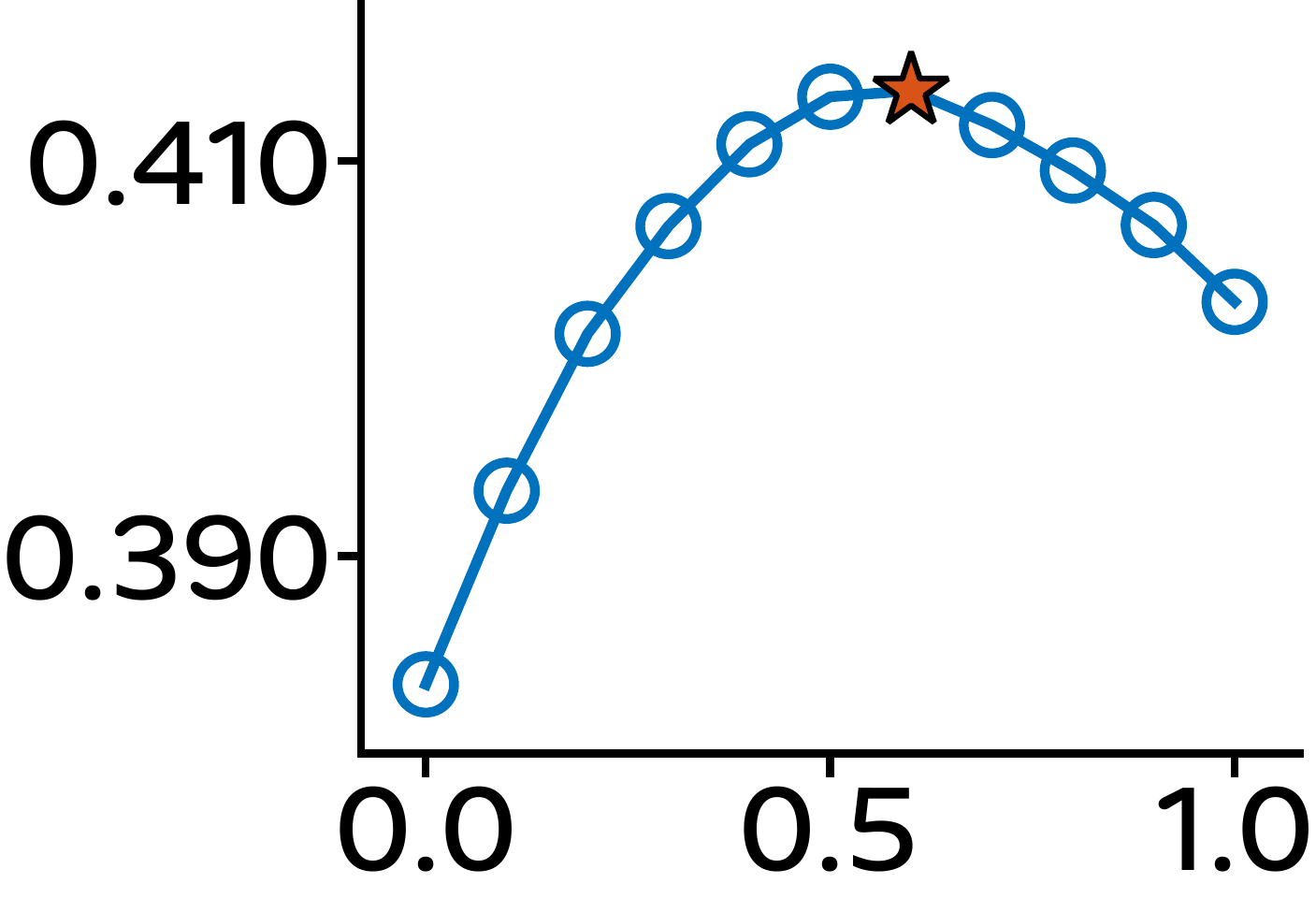}
        \captionsetup{margin={6mm,0mm}}
        \caption{Arts}
    \end{subfigure}
    \begin{subfigure}[c]{0.31\linewidth}
        \includegraphics[width=\linewidth]{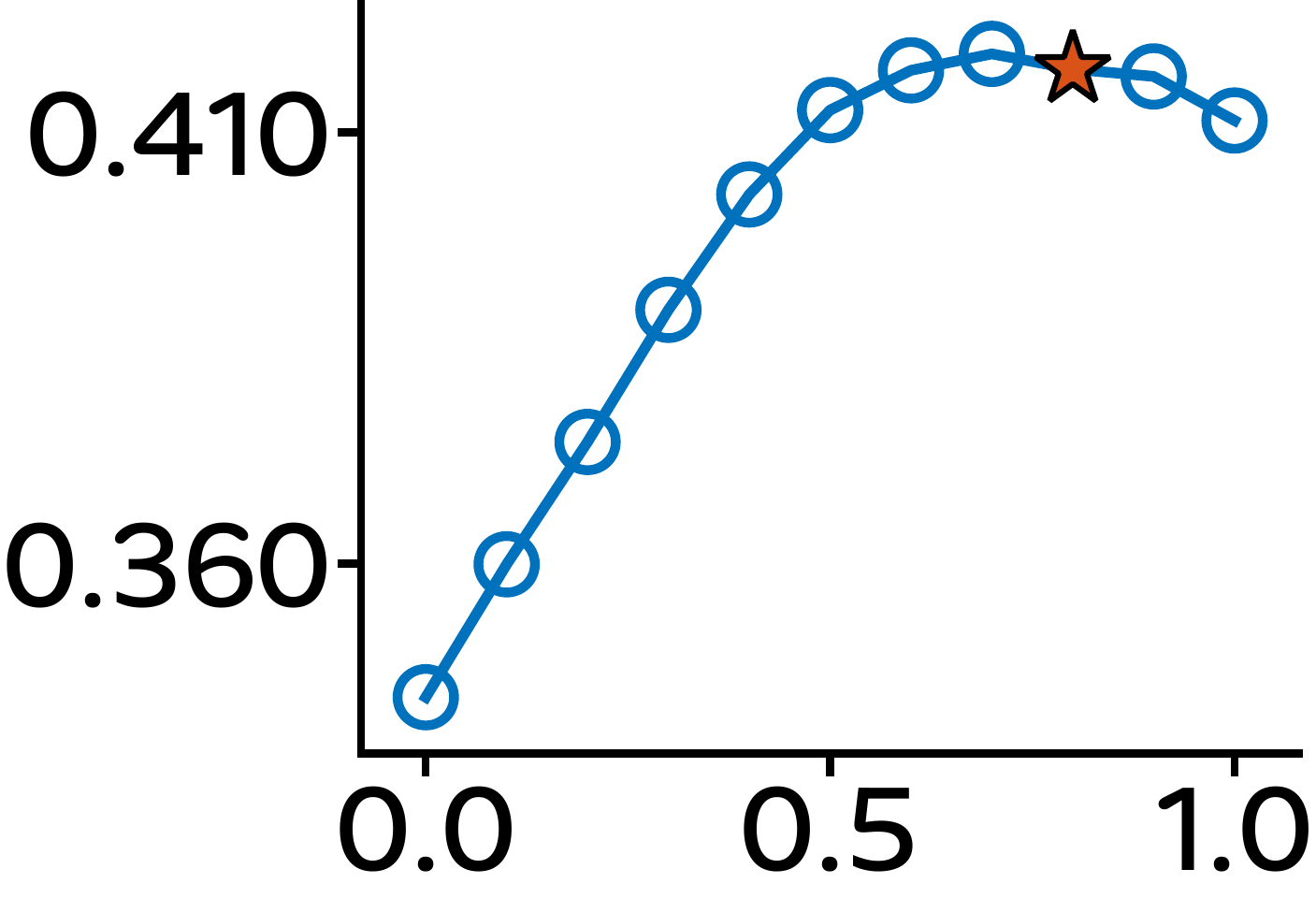}
        \captionsetup{margin={6mm,0mm}}
        \caption{Movies}
    \end{subfigure}
    \begin{subfigure}[c]{0.31\linewidth}
        \includegraphics[width=\linewidth]{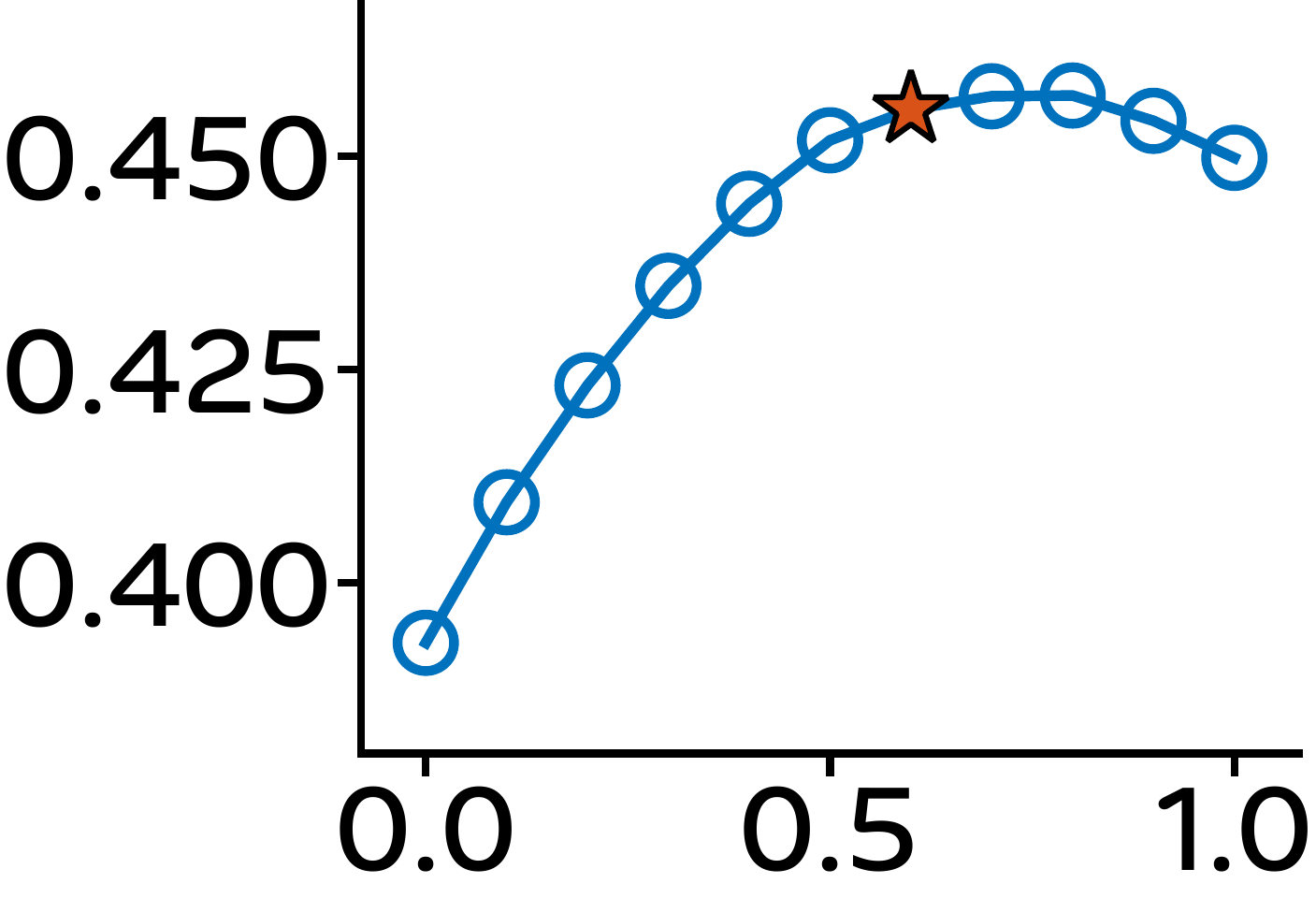}
        \captionsetup{margin={6mm,0mm}}
        \caption{Video}
    \end{subfigure}
    \caption{Sensitivity analysis of the fusion weight $\beta$, with stars indicating validation-selected values. Performance peaks at intermediate values, highlighting the importance of balancing the collaborative relation score with the local score.}
    \label{fig:exp:beta_sensitivity}
    \Description{Three line charts plot HR@10 against the fusion weight from zero to one. Stars mark the validation-selected weights on each curve. The curves rise and then decline after reaching intermediate maxima.}
\end{figure}

Figure~\ref{fig:exp:beta_sensitivity} shows the effect of the fusion weight $\beta$, which controls the contribution of collaborative memory. Performance initially improves as $\beta$ increases, peaks at an intermediate value, and declines when $\beta$ becomes too large. A small $\beta$ underuses collaborative information, whereas an excessively large $\beta$ weakens the contribution of local preference signals.
{\color{black}
The validation-selected values range from 0.4 to 0.8, with a median of 0.6, showing the importance of balancing the two scores.
}

\subsection{Adaptation to Evolving Client Behavior}
\label{sec:exp:adaption}
Our main experiments assume fixed client histories, whereas new interactions may make previously retrieved client-specific memories outdated. We therefore examine how well \method adapts to evolving client behavior.
\subsubsection{Performance under different refresh strategies}
\begin{figure}[t]
    \centering
    \includegraphics[width=0.85\linewidth,trim=0 8bp 0 7bp,clip]{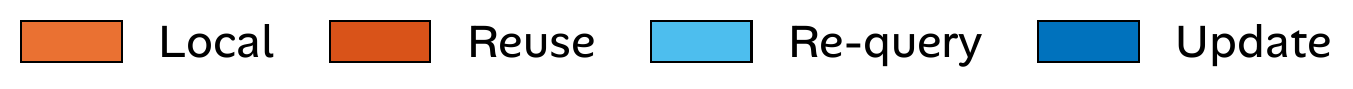}
    \\[2pt]
    \begin{subfigure}[c]{0.0275\linewidth}
        \raisebox{4.5mm}{\includegraphics[width=\linewidth]{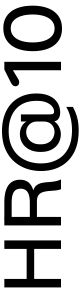}}
    \end{subfigure}
    \begin{subfigure}[c]{0.313\linewidth}
        \includegraphics[width=\linewidth]{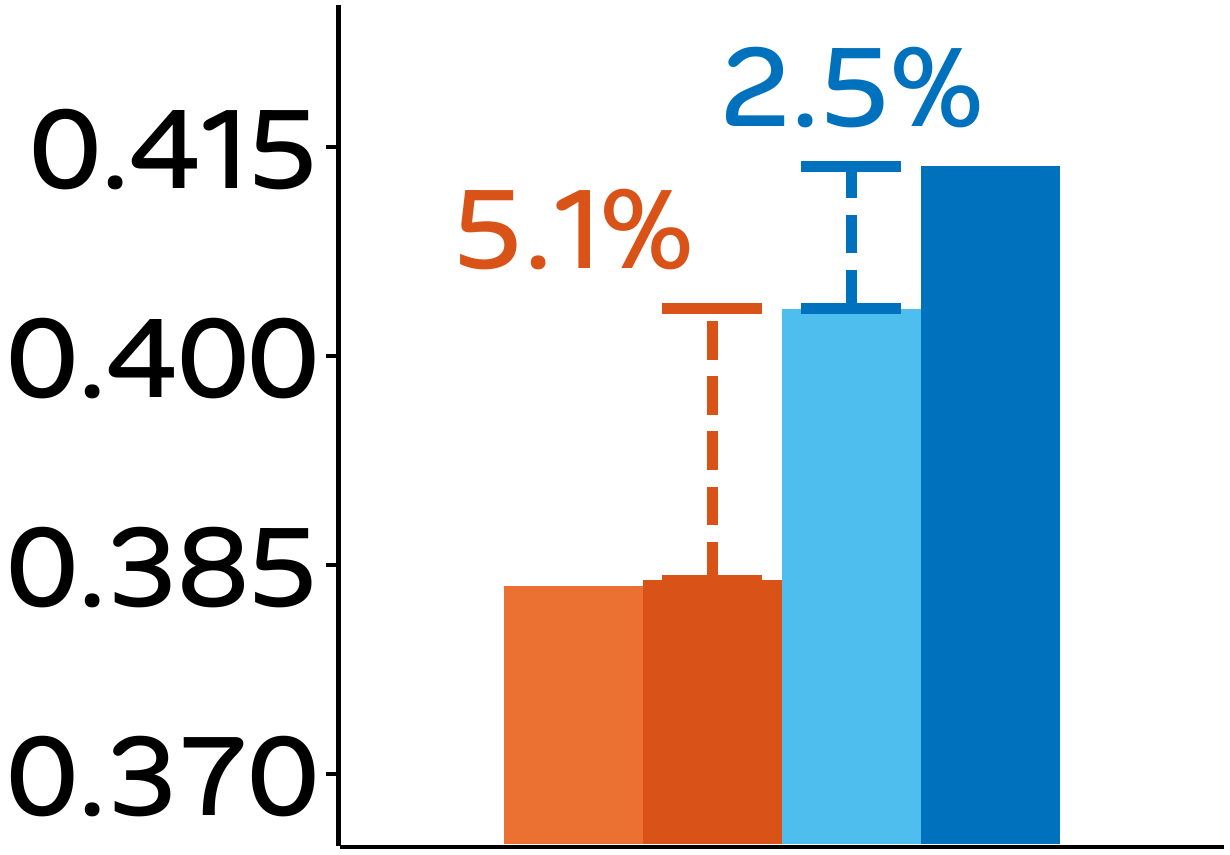}
        \captionsetup{margin={6mm,0mm}}
        \caption{Arts}
    \end{subfigure}
    \begin{subfigure}[c]{0.313\linewidth}
        \includegraphics[width=\linewidth]{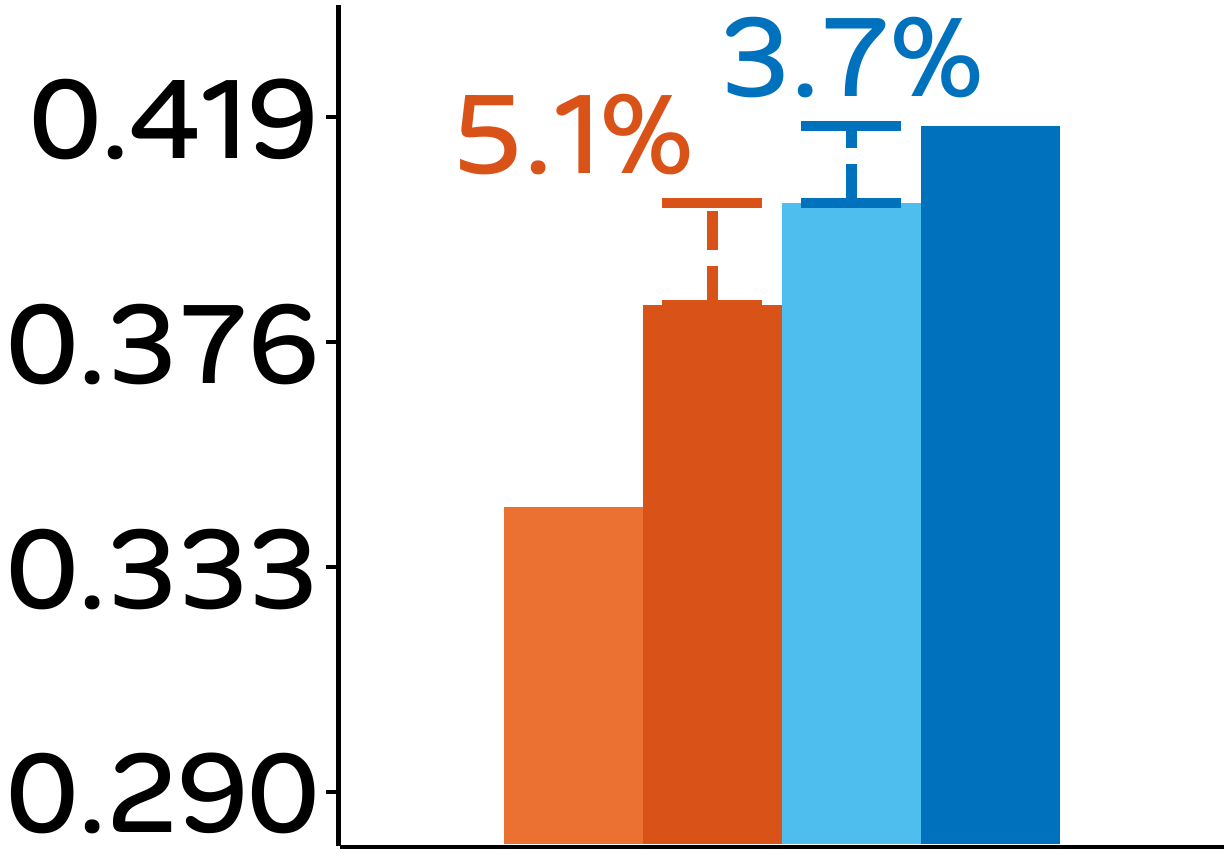}
        \captionsetup{margin={6mm,0mm}}
        \caption{Movies}
    \end{subfigure}
    \begin{subfigure}[c]{0.313\linewidth}
        \includegraphics[width=\linewidth]{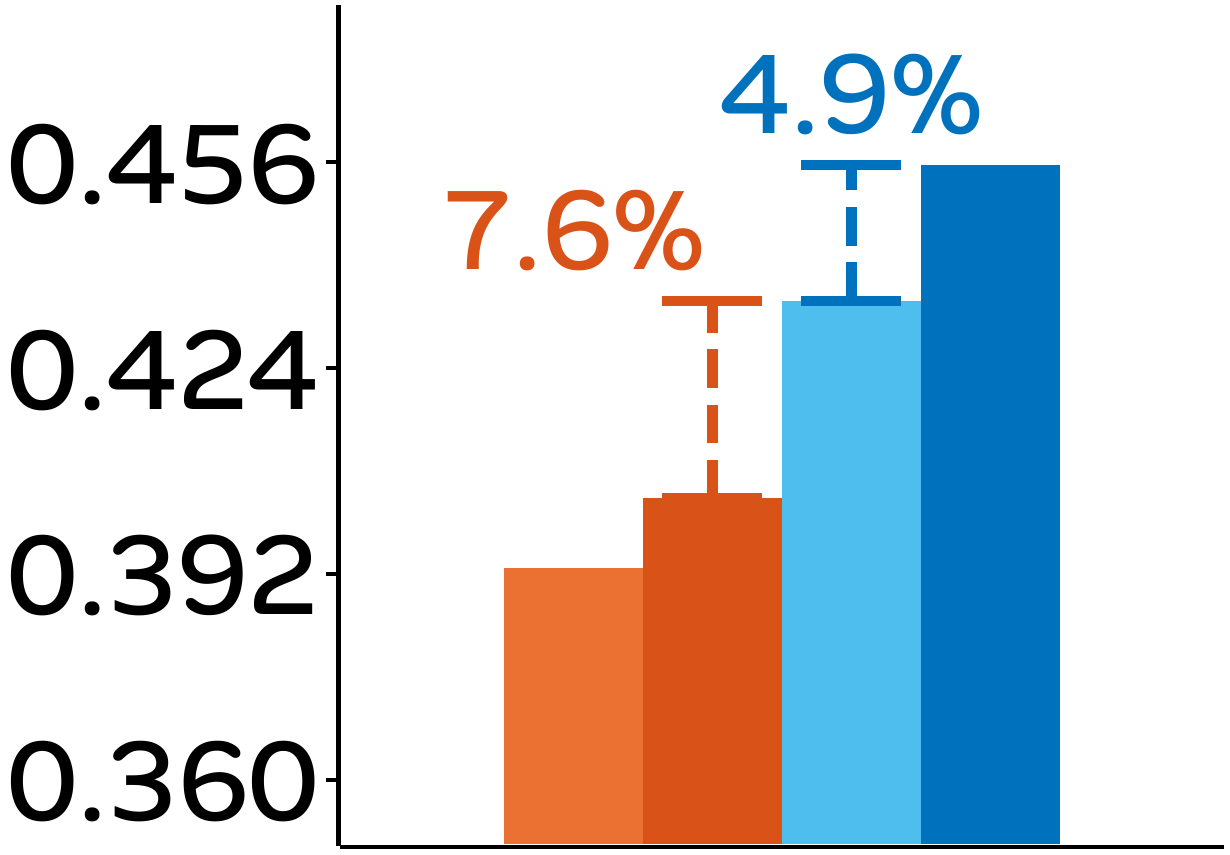}
        \captionsetup{margin={6mm,0mm}}
        \caption{Video}
    \end{subfigure}
    \caption{    
    Recommendation performance under different memory refresh strategies. Re-querying with accumulated interactions improves performance without updating the memory bank.
    }
    \label{fig:temporal_transition}
    \Description{Three bar charts compare Local, Reuse, Re-query, and Update on Arts, Movies, and Video. Annotations show the percentage gains from Reuse to Re-query and from Re-query to Update.}
\end{figure}

We chronologically split each client’s interactions into old and new portions at 7:3 ratio and construct the initial memory bank using only the old interactions. 
At evaluation time, all variants use the accumulated local history. 
{\color{black}
\texttt{Local} uses no collaborative memory, \texttt{Reuse} retains the initially retrieved client memory, \texttt{Re-query} retrieves a new client memory using the updated client query, and \texttt{Update} rebuilds the memory bank before retrieval.
Figure~\ref{fig:temporal_transition} shows that \texttt{Reuse} improves over \texttt{Local}, indicating that the initial memory remains useful, while \texttt{Re-query} further improves performance without reconstructing the memory bank.
\texttt{Update} achieves the best performance at additional rebuilding cost, indicating a trade-off between adaptation and system overhead.
}

\subsubsection{Effect of behavioral changes on memory retrieval}
\begin{figure}[t]
    \centering
    \includegraphics[width=0.95\linewidth,trim=0 11bp 0 11bp,clip]{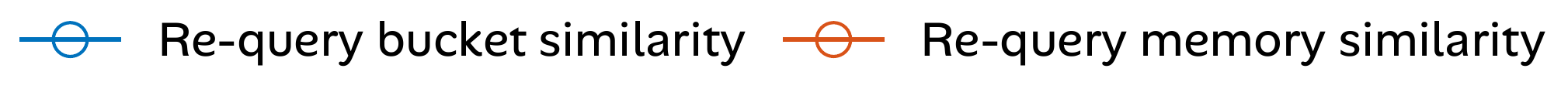}
    \\[0pt]
    \begin{subfigure}[c]{0.032\linewidth}
        \raisebox{6mm}{\includegraphics[width=\linewidth]{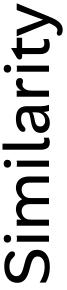}}
    \end{subfigure}
    \begin{subfigure}[c]{0.31\linewidth}
        \includegraphics[width=\linewidth]{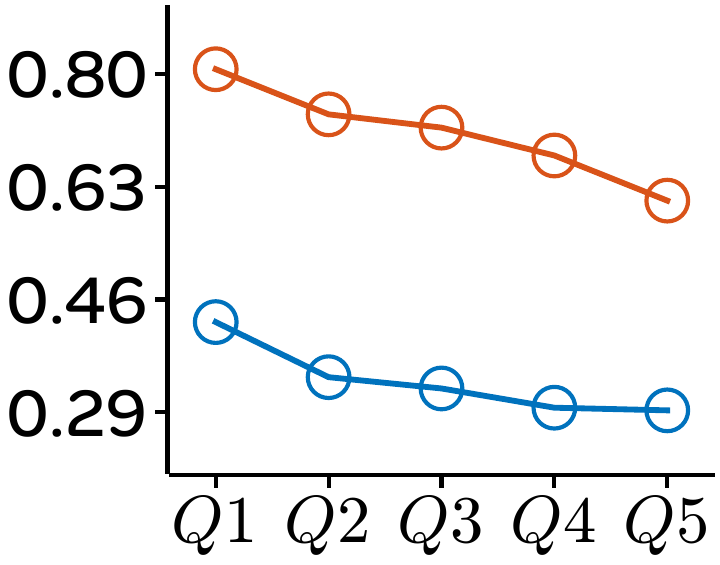}
        \captionsetup{margin={6mm,0mm}}
        \caption{Arts}
    \end{subfigure}
    \begin{subfigure}[c]{0.31\linewidth}
        \includegraphics[width=\linewidth]{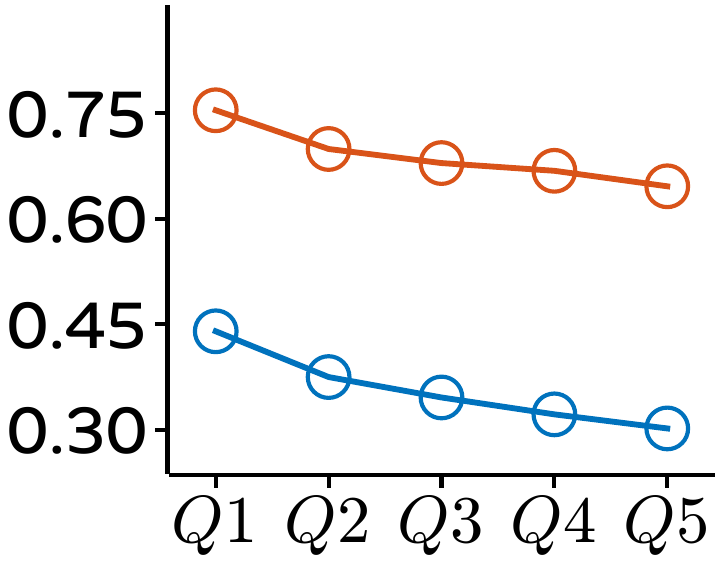}
        \captionsetup{margin={6mm,0mm}}
        \caption{Movies}
    \end{subfigure}
    \begin{subfigure}[c]{0.31\linewidth}
        \includegraphics[width=\linewidth]{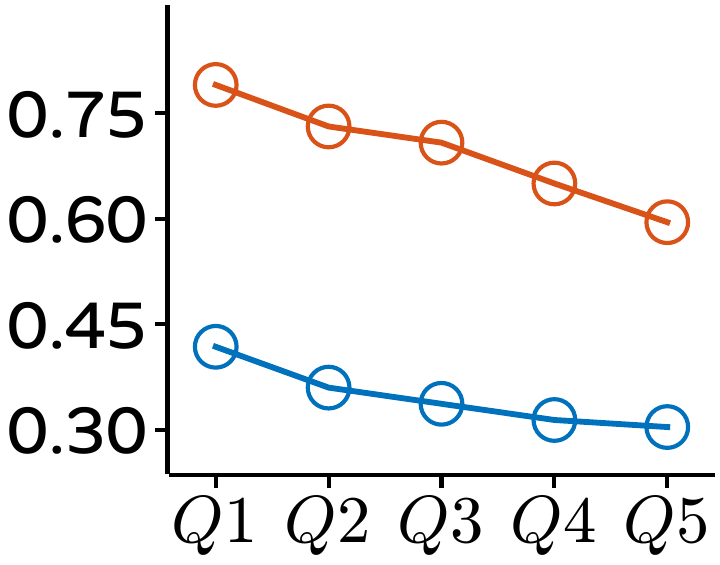}
        \captionsetup{margin={6mm,0mm}}
        \caption{Video}
    \end{subfigure}
    \caption{
    Similarity of retrieved buckets and client memories across behavioral change levels. As behavioral changes increase, re-query retrieves increasingly different buckets, resulting in more distinct client memories.
    }
    \label{fig:behavior_drift}
    \Description{Three line charts plot bucket-set similarity and memory similarity over five groups of increasing behavioral change. Both curves decline from Q1 to Q5, and memory similarity remains above bucket-set similarity.}
\end{figure}

We divide clients into five groups based on the weighted Jaccard similarity between the normalized interaction histograms of their old (70\%) and new (30\%) histories. Q1 contains clients with the most similar behaviors before and after evolution, while Q5 contains those with the largest behavioral changes. Re-query bucket similarity measures the Jaccard similarity between the bucket sets retrieved before and after re-querying, while Re-query memory similarity measures the cosine similarity between the corresponding client memories.
Figure~\ref{fig:behavior_drift} shows that both similarities decrease from Q1 to Q5 across all datasets. As client behavior changes more substantially, Re-query selects different buckets and constructs a more distinct client memory. This confirms that Re-query adapts collaborative information to evolving client behavior.

\begin{figure}[!t]
    \centering
    \includegraphics[width=0.9\linewidth,trim=0 17bp 0 9bp,clip]{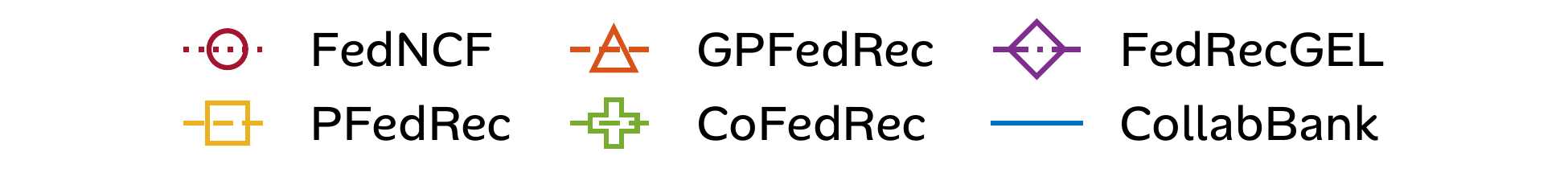}
    \\[-1mm]
    \begin{subfigure}[t]{0.03\linewidth}
        \raisebox{3mm}{\includegraphics[width=\linewidth]{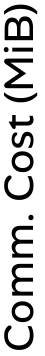}}
    \end{subfigure}
    \begin{subfigure}[t]{0.313\linewidth}
        \includegraphics[width=\linewidth]{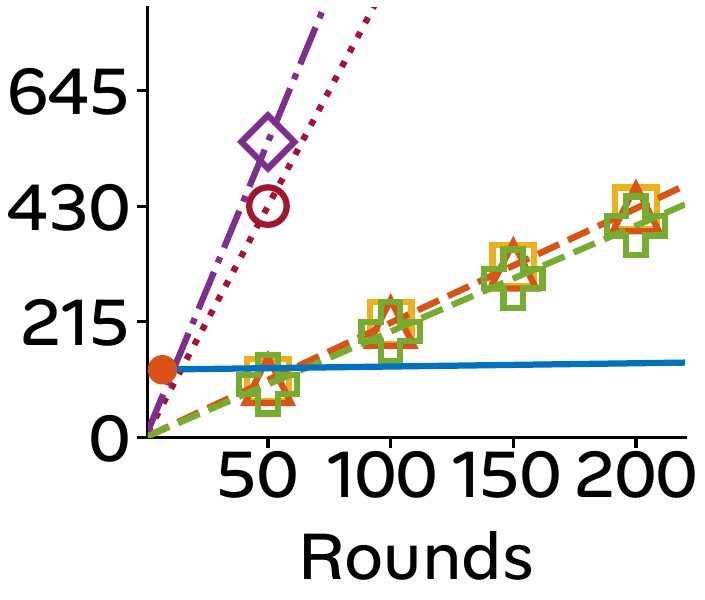}
        \captionsetup{margin={4mm,0mm}}
        \caption{Arts}
    \end{subfigure}
    \begin{subfigure}[t]{0.313\linewidth}
        \includegraphics[width=\linewidth]{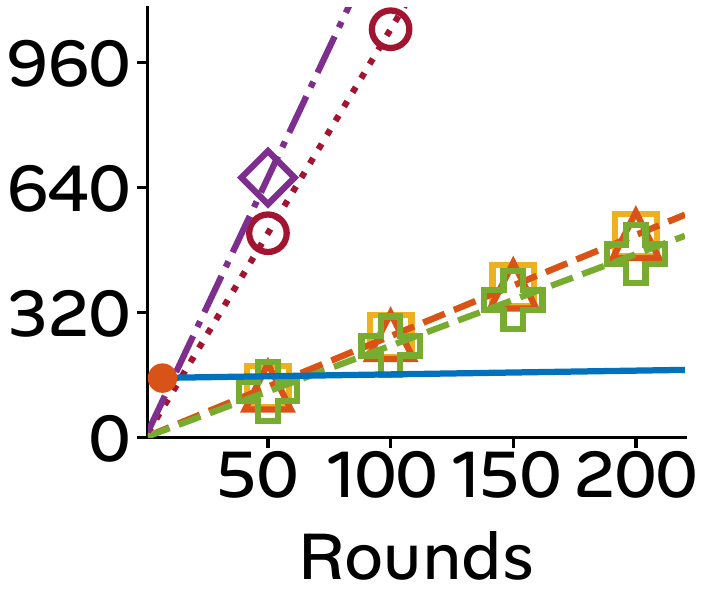}
        \captionsetup{margin={4mm,0mm}}
        \caption{Movies}
    \end{subfigure}
    \begin{subfigure}[t]{0.313\linewidth}
        \includegraphics[width=\linewidth]{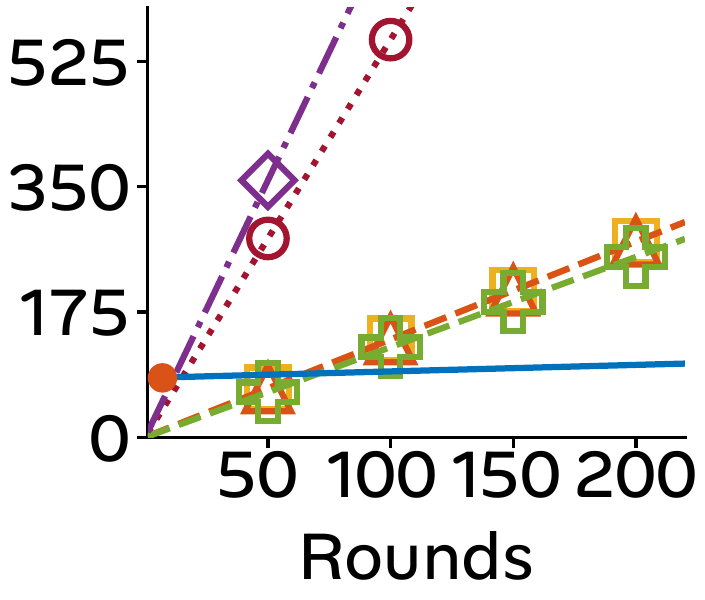}
        \captionsetup{margin={4mm,0mm}}
        \caption{Video}
    \end{subfigure}
    \caption{
    Cumulative per-client communication cost across rounds, including one-time initialization. \method requires only a single collaborative exchange (orange dot), while its compact payload keeps the cumulative cost low, with only a gradual increase under repeated exchanges.
    }
    \label{fig:exp:communication_cost}
    \Description{Three line charts plot cumulative communication cost in MiB over 200 rounds. Curves for five federated baselines are compared with the much flatter CollabBank curve; an orange point marks its first collaborative exchange.}
\end{figure}

\subsection{Communication Efficiency}
\label{sec:exp:communication}

We compare the cumulative per-client communication cost of \method with the baselines across rounds in Figure~\ref{fig:exp:communication_cost}, where FedCIA is omitted due to its $O(|\Items|^2)$ communication.
For \method, we accumulate one query-response exchange per round after the one-time initialization, although only a single collaborative exchange is required for recommendation.
Overall, its compact per-exchange payload causes only a small but nonzero increase in cumulative communication, keeping the cost substantially below iterative federated recommendation methods across rounds.

\section{Conclusion}
In this work, we propose \method, a memory-based approach that reformulates federated collaboration as retrieval over compact server-side memories. 
\method organizes semantically related items using pretrained language-model embeddings and distills collaborative evidence from clients with similar behavioral patterns into separate memories. 
For each target client, the server retrieves and combines only query-relevant memories, enabling client-specific knowledge transfer while reducing interference from irrelevant collaborative information. 
Moreover, once the memory bank has been constructed, each client can obtain complementary collaborative knowledge through a single communication round. 
Extensive experiments show that \method achieves strong recommendation accuracy compared with existing methods, while validating the effectiveness of its semantic grouping and collaborative memory design. 
These results suggest that retrieval-based memory offers an effective and communication-efficient direction for personalized collaboration in heterogeneous federated recommendation systems.

\section*{Ethical Considerations}

A common design in existing federated recommendation is to keep raw interaction histories local while exchanging client-specific updates or representations derived from local interactions~\cite{ZhangL0YZZY23,ZhangL0ZYY24}.
Prior studies have shown that such messages may still leak interaction or preference information~\cite{HuS24}.
\method represents each client's behavior as a compact sketch over shared semantic clusters and does not directly transmit item identities or raw interaction histories.
However, because the sketch preserves cluster-level behavior and client similarity for collaborative retrieval, it may reveal coarse-grained semantic preferences and behavioral similarities across clients.
Thus, the current design avoids directly exposing fine-grained item-level behavior, but coarse-grained preferences may remain observable.
Deployments with stronger privacy requirements could apply local differential privacy to perturb client sketches before transmission or employ secure computation for similarity-based retrieval without directly exposing individual sketches to the server.

\begin{acks}
This work was supported by Korea Internet \& Security Agency (KISA) grant funded by the Korea government (PIPC) (No.RS-2026-25526342).
This work was supported by the Institute of Information \& Communications Technology Planning \& Evaluation (IITP) - Innovative Human Resource Development for Local Intellectualization program grant funded by the Korea government (MSIT) (IITP-2027-RS-2022-00156360) and 
the Convergence security core talent training business support program (IITP-2027-RS-2024-00426853).
\end{acks}

\balance
\bibliographystyle{ACM-Reference-Format}
\bibliography{myref}
\end{document}